\documentclass[aps, prb, twocolumn, amsmath, amssymb, showpacs,
superscriptaddress]{revtex4-1}
\usepackage{graphicx}
\usepackage{psfrag}
\usepackage{dcolumn} 
\usepackage{bm} 
\usepackage{hyperref} 
\usepackage{color}
\usepackage{url}

\graphicspath{{./pictures/}}

\begin{document}

\title{Low-temperature structure of \texorpdfstring{$\xi'$}{xi'}-Al--Pd--Mn
  optimized by \textit{ab initio} methods}

\author{Benjamin Frigan}
\email{benjamin.frigan@itap.uni-stuttgart.de}
\affiliation{Institut f\"{u}r Theoretische und Angewandte Physik,
  Universit\"{a}t Stuttgart, 70550 Stuttgart, Germany}
\author{Alejandro Santana}
\affiliation{Grupo de sistemas complejos, Universidad Antonio Nari\~{n}o,
  Bogot\'{a}, Colombia}
\author{Michael Engel}
\affiliation{Department of Chemical Engineering, University of Michigan, Ann
  Arbor, MI 48109, USA}
\author{Daniel Schopf}
\affiliation{Institut f\"{u}r Theoretische und Angewandte Physik,
  Universit\"{a}t Stuttgart, 70550 Stuttgart, Germany}
\author{Hans-Rainer Trebin}
\affiliation{Institut f\"{u}r Theoretische und Angewandte Physik,
  Universit\"{a}t Stuttgart, 70550 Stuttgart, Germany}
\author{Marek Mihalkovi\v{c}}
\affiliation{Institute of Physics, Slovak Academy of Sciences, 84511
  Bratislava, Slovakia}

\date{\today}

\begin{abstract}
We have studied and resolved occupancy correlations in the existing
average structure model of the complex metallic alloy $\xi'$-Al--Pd--Mn [Boudard
\textit{et al.}, Phil.~Mag.~A \textbf{74} 939 (1996)], which has approximately
320 atoms in the unit cell and many fractionally occupied sites.
Model variants were constructed systematically in a tiling-decoration approach
and subjected to simulated annealing by use of both density functional theory
and molecular dynamics with empirical potentials. To obtain a measure for
thermodynamic stability, we reproduce the Al--Pd--Mn phase diagram at $T$=0~K,
and derive an enthalpy of formation for each structure. 
Our optimal structure resolves a cloud of fractionally occupied sites in
pseudo-Mackay clusters. In particular, we demonstrate the presence of
rotational degrees of freedom of an Al$_{9}$ inner shell, which is caged
within two icosahedrally symmetric outer shells Al$_{30}$ and
Pd$_{12}$. Outside these clusters, the chemical ordering on a chain of three
nearby sites surprisingly breaks the inversion symmetry of the surrounding
structure, and couples to an Al/vacancy site nearby.
Our refined tiling-decoration model applies to any structure within the
$\varepsilon$-phases family, including the metastable decagonal
quasicrystalline phase.
\end{abstract}


\maketitle

\section{Introduction}
\label{sec:intro}
The ternary Al--Pd--Mn system is of great interest for the study of
physical properties of quasicrystals and their approximants. Its phase diagram
contains both a stable icosahedral phase and a stable decagonal phase in the
vicinity of various other crystalline approximants, many of which can be grown
as single crystals with high perfection. Of particular interest is the series
$\xi'$, $\xi_{1}'$, $\xi_{2}'$, \dots of orthorhombic
approximants~\cite{heggen_structural_2008} of the icosahedral quasicrystalline
alloy $i$-Al--Pd--Mn, which exists in the Al-rich corner of the phase
diagram.\cite{raghavan_2009} An alternative notation for these phases is
$\varepsilon_l$, $l=6, 16, 28, \dots$, where $l$ denotes the strong $(0,0,l)$
diffraction spot for the interplanar spacing. $\xi_{2}'$  ($\varepsilon_{28}$)
is frequently named $\Psi$-phase. All these phases have a common periodicity
of 16~\AA\ along $(0,1,0)$ and can be described as two-dimensional tilings
perpendicular to that direction.\cite{heggen_structural_2008} For the
$\xi'$-phase ($\varepsilon_{6}$) the tiling consists of staggered flattened
hexagon tiles, where occasionally \emph{phason defects} can show up composed
of an nonagon and pentagon tile.\cite{klein_phason_1996} There is a similar
structure where the hexagons are all parallel, denoted $\xi$. In the
$\xi_{n}'$-phases, the phason defects are arranged in rows, called
\emph{phason planes} and squeezed in between $n-1$ rows of hexagons forming a
layered superstructure. 

The most recent experimental structure refinement of $\xi'$ was conducted in
1996 by Boudard {\it et al.}\cite{boudard_structure_1996} The close
relationship between $\xi$ and $\xi'$ was studied by Klein {\it et al.} via
high resolution electron microscopy.\cite{klein_phason_1996} So far, no
structure refinement of the $\xi_n'$-phases has been reported in the 
literature, but the tiling representation allows to extend structure models
for $\xi$ and $\xi'$ to structure models of $\xi_n'$. Important building
blocks of all structures are pseudo-Mackay icosahedral clusters (PMI) which
form columns parallel to $(0,1,0)$ and when projected are the vertices of the
tiling representation discussed above. The PMI shares the outer shells
(Pd$_{12}$ icosahedron and Al$_{30}$ icosidodecahedron) with the conventional
Mackay icosahedron (MI) cluster, while the inner shell, an Al$_{12}$
icosahedron in the MI, is replaced by a less symmetric shell of 8--10 Al
atoms. While the positions of the outer shells are predicted with high
accuracy from diffraction data, the positions of the inner Al atoms are less
well defined and have large uncertainties: in experiment, the inner sites are
characterized by mixed and partial occupancies;\cite{boudard_structure_1996}
when projected from a hyper-space model they correspond to positions close to
the boundary of atomic surfaces,\cite{beraha_correlated_1997} which suggests
that they might be intrinsically unstable.

Interest in the $\xi_n'$-phases has spurred when Klein {\it et
  al.}\cite{Klein_1999} discovered a special texture around a partial 
dislocation of Burgers vector 1.83~\AA.\footnote{The $\xi_n'$-phases have
  large lattice constants, which -- due to too high strain energies -- prevent
  the existence of integer dislocations and admit only partial ones.}  On the
tiling level,\cite{engel_unified_2005} the dislocation results in the
insertion of six phason planes of width $\approx$ 171~\AA, which means it can
alternatively be viewed as a dislocation in the layered superstructure. The
dual nature of the dislocation -- on the atomic level and on the tiling level
-- and the resulting hierarchy in length scales lead to the adoption of the term
\emph{metadislocation} for this special class of dislocations.\cite{Klein_1999}
Metadislocations are known to be important for the plasticity of the
$\xi_n'$-phases,\cite{Klein_1999,Feuerbacher_2001,Heggen_metareac_2005} but
so far little is known about the atomistic structure of the metadislocation
core and the details of the metadislocation motion. A refined atomistic
structure model of the underlying crystal phases as achieved in this work can
form the basis of a better understanding of these two aspects.

In this contribution we report a structural refinement of the Boudard model of
$\xi'$ via numerical optimization. Simulated annealing with molecular dynamics
and empirical potentials, and density functional theory are used to relax the
structure into its energy minimum. We derive an enthalpy of formation for each
model and compare it with the enthalpy of other phases coexisting in the
Al--Pd--Mn system. This yields a direct measure for thermodynamic stability. We
identify two main sources of disorder that plagued a direct structural
determination of the structure from the diffraction data. One type of disorder
is related to the symmetry-broken inner shell of the PMI cluster. The
symmetry-breaking of this shell, along with extra space due to the vacated
sites, leads to a low-temperature state described by an ensemble of
configurations, instead of the unique icosahedral one. Individual
representatives of this ensemble are {\it a priori} not accessible to
experimental diffraction refinements, which only captures the average
structure. We discuss how such an average structure may even become
ill-defined due to missing strong intra-cluster and inter-cluster occupancy
correlations. The second source of disorder mixes vacancies, Pd and Al atoms
in the interstitial parts of the structure, not covered by any PMI cluster. 

The symmetry-broken PMI inner-shell configurations may acquire special
importance due to their potential impact on the long-range ordering in
quasicrystals. For it is exactly these atoms that define boundaries of
polyhedral atomic surfaces defining perfect quasicrystal structures in
six-dimensional hyperspace.\cite{Katz_1993} While there is probably no hope to
accurately determine such individual configurations {\it in situ} in the
quasicrystal phase, crystal structures like the $\xi'$-phase studied here
present a unique opportunity to study these configurations without the need of
modelling a quasiperiodic quasicrystalline structure. 

The paper is organized as follows: After a description of the initial model
and simulation methods in Sec.~\ref{sec:boudard-model} and
\ref{sec:comp-methods}, respectively, we present details of our results in
Sec.~\ref{sec:results}: first, a low-temperature Al--Pd--Mn phase diagram is
presented in Sec.~\ref{sec:alpdmn-phase-diagram}, which forms the basis for
our stability evaluations. The optimization of the $\xi'$-phase and the
rotational degrees of freedom of the inner PMI shells are discussed in
Sec.~\ref{sec:Optimization} and Sec.~\ref{sec:ab-intio-MD}. Finally, in
Sec.~\ref{sec:discussion} a comparison is made with the isostructural
Al$_{3}$Pd phase, and first structure models for the phason defects are
presented. We conclude with a short summary in Sec.~\ref{sec:summary}.

\section{Average Structure of \texorpdfstring{$\xi'$}{xi'}: The Boudard Model}
\label{sec:boudard-model}
The space group of the $\xi'$-phase was determined by Boudard \textit{et al.}
\cite{boudard_structure_1996} as \textit{Pnma} (No.~62).\footnote{The number
  in parentheses specifies the space group according to the International
  Tables for Crystallography, Volume A: Space-Group Symmetry} The unit cell
contains about 320 atoms with a composition around 
Al$_{73.5}$Pd$_{22.4}$Mn$_{4.1}$. The lattice parameters are $a=23.54$~\AA,
$b=16.57$~\AA\ and $c=12.34$~\AA. The structure exhibits a large amount of
disorder. Only 19 of 48 atomic sites are fully occupied, whereas 15 sites
possess occupancy factors of 0.5 or less.
\begin{figure}[!h]
  \begin{minipage}[c]{0.49\linewidth}
    \centering
    \includegraphics[scale=0.11]{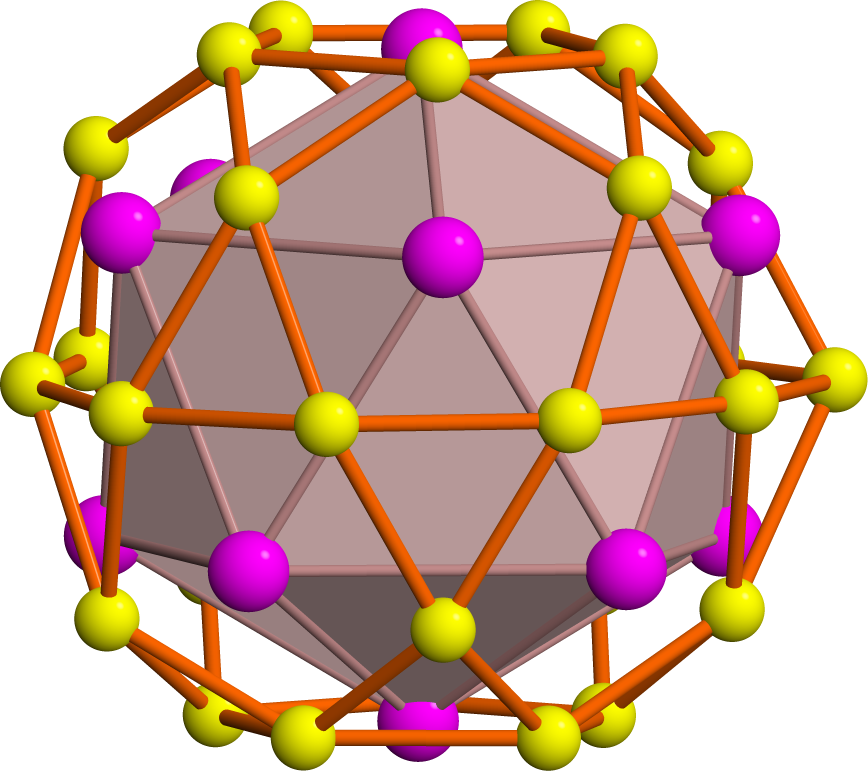}
  \end{minipage}
  \begin{minipage}[c]{0.49\linewidth}
    \centering
    \includegraphics[scale=0.135]{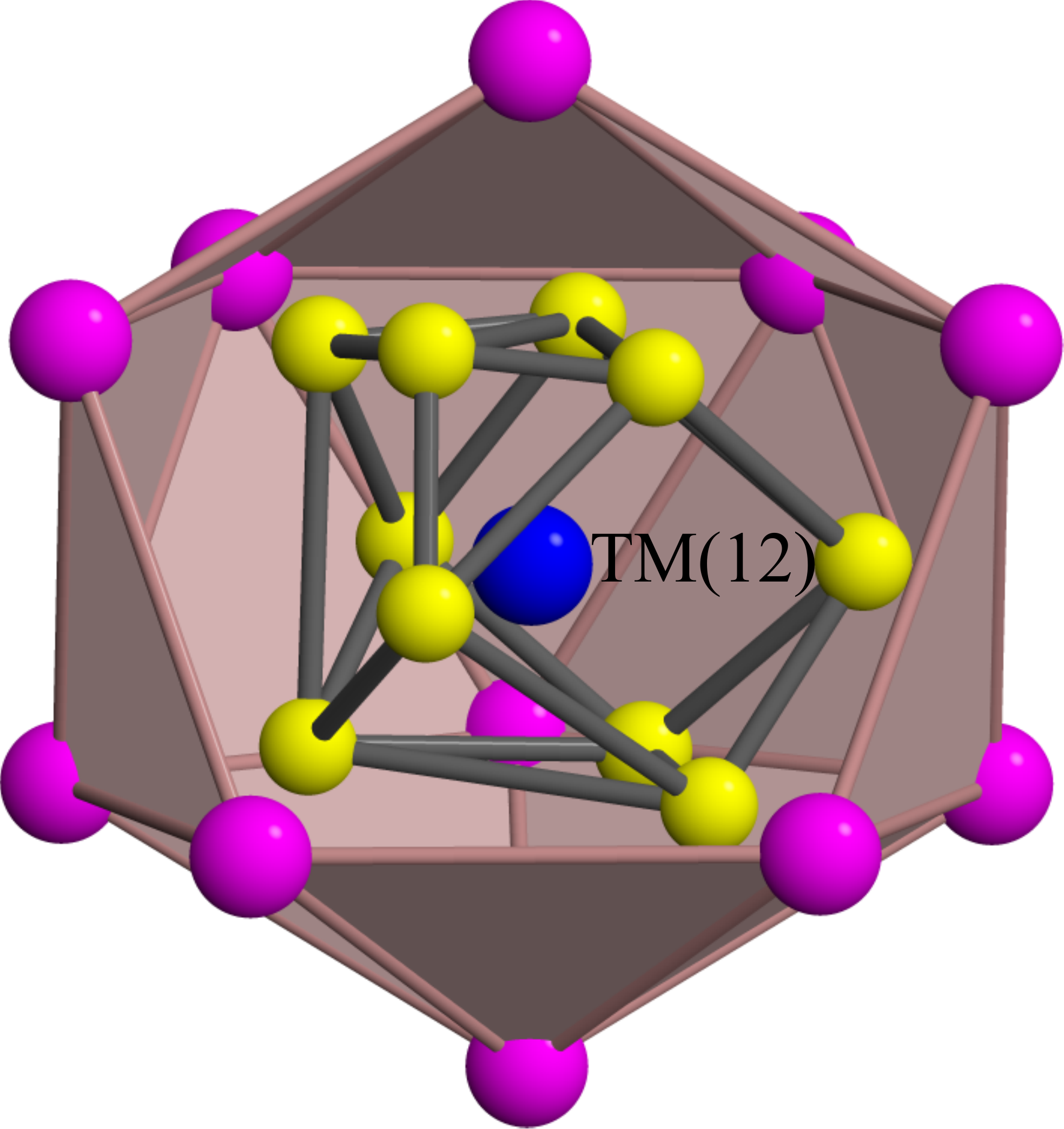}
  \end{minipage}

\vspace{-0.4cm}
  \centering (a)\\

  \vspace{0.3cm}
  \begin{minipage}[c]{0.49\linewidth}
    \centering
    \includegraphics[scale=0.11]{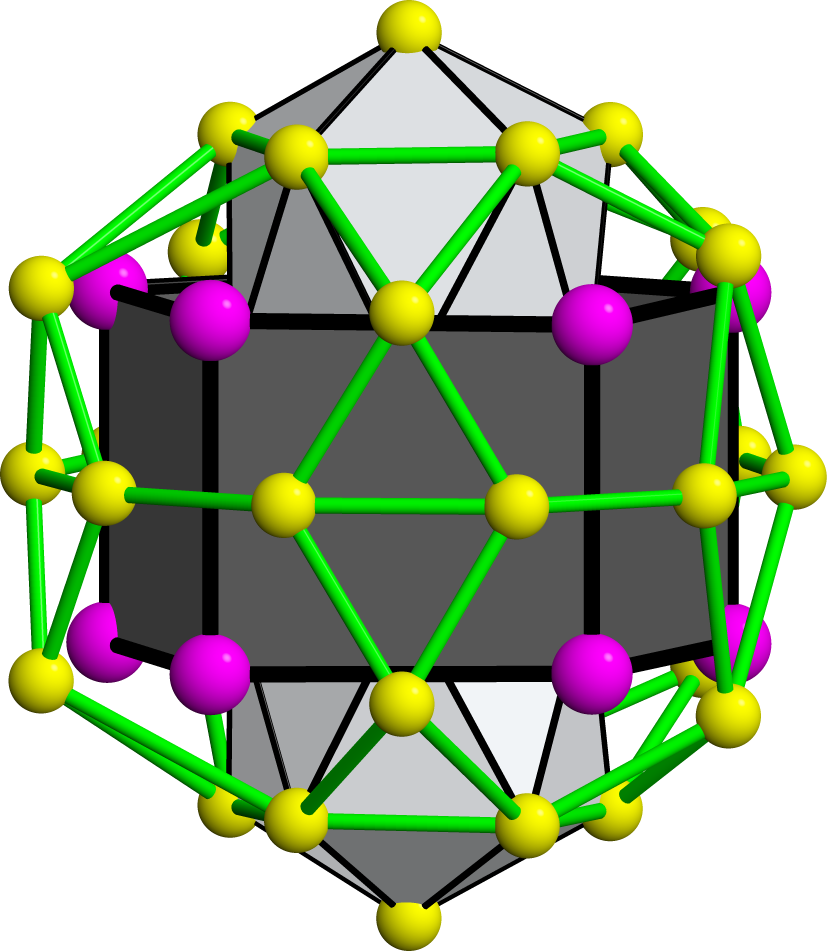}
  \end{minipage}
  \begin{minipage}[c]{0.49\linewidth}
    \centering
    \includegraphics[scale=0.135]{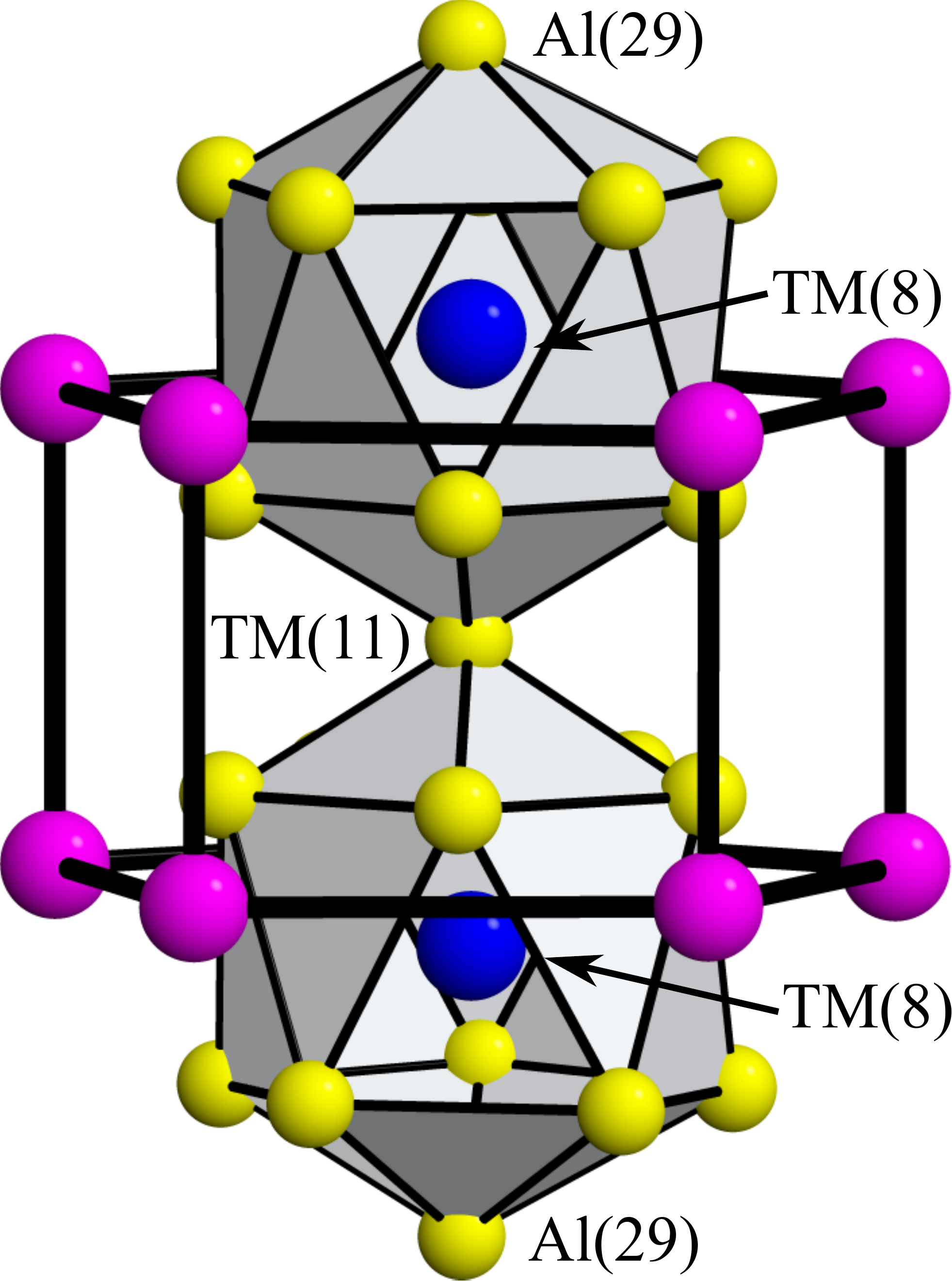}
  \end{minipage}

\vspace{-0.4cm}
  \centering (b)\\[0.2cm]
  \fbox{
  \includegraphics[scale=0.05]{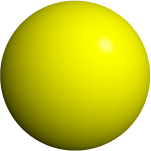}\hspace{0.2mm}Al \ \ 
  \includegraphics[scale=0.05]{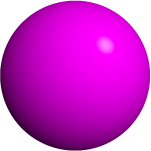}\hspace{0.2mm}Pd \ \
  \includegraphics[scale=0.05]{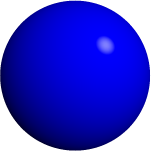}\hspace{0.2mm}Mn}
  \caption{(Color online) Atomic clusters in $\xi'$.
    (a) PMI cluster consisting of a centered Mn atom, a symmetry-broken inner Al
    shell, a Pd icosahedron, and an Al icosidodecahedron. (b) LBPP cluster
    composed of two Mn-centered Al icosahedra which are capped within a Pd
    pentagonal prism, and an Al icosidodecahedron-type shell. Atomic
    coordinates and labeling is taken from
    Ref.~\onlinecite{boudard_structure_1996}. Note that in these pictures all
    atomic sites are fully occupied.}  
  \label{fig:xi_clusters}
\end{figure}

The average structure of $\xi'$ consists of two types of atomic
clusters, the PMI and the so-called large bicapped pentagonal prism (LBPP). 
The clusters are illustrated in Fig.~\ref{fig:xi_clusters}. The PMI cluster is
formed by a symmetry-broken (relative to the approximate icosahedral symmetry
of the outer shell) inner shell, which accommodates 8--10 Al atoms. Most of
the atoms form short (2.35--2.5~\AA), presumably strong bonds with the
central Mn atom. This inner shell is encaged by the nearly icosahedral second
shell, forming apparently favourable Al-Pd patterns with Pd atoms. The second
shell is composed of a large Pd$_{12}$ icosahedron with a radius of 4.44~\AA,
and of an Al$_{30}$ icosidodecahedron with an approximate radius of
4.9~\AA. Both Al and Pd subshells of the second shell have only about
0.14~\AA\ radial deformation. The significance of this cluster is clearly
evidenced by the 1.5~\AA\ gap separating the second shell from the outer
atomic structure.

The PMI clusters cover almost 90\% of the structure. All remaining atoms can
be ascribed to a LBPP cluster. Despite its nearly-perfect pentagonal symmetry,
this cluster also consists of two nearly spherical shells, that are well
separated from each other, and also from the outer structure. Here, the
first-neighbor distances to the central atom range from 2.6 to 2.8~\AA\ for
the inner Al$_{10}$Mn$_{2}$ shell. The outer Pd$_{10}$Al$_{32}$ shell
has about the same radii as the second shell of the PMI cluster (4.6~\AA\ for
Pd atoms, 5.0~\AA\ for Al atoms), with a bigger radial distortion
(about 0.3~\AA) for the Al subshell. It is still well separated from the outer
shell, which starts at approximately 6.5~\AA. Only 9~atoms of the LBPP cluster
are not shared with any adjacent PMI.
\begin{figure}[!h]
  \centering
  \includegraphics[width=0.8\linewidth]{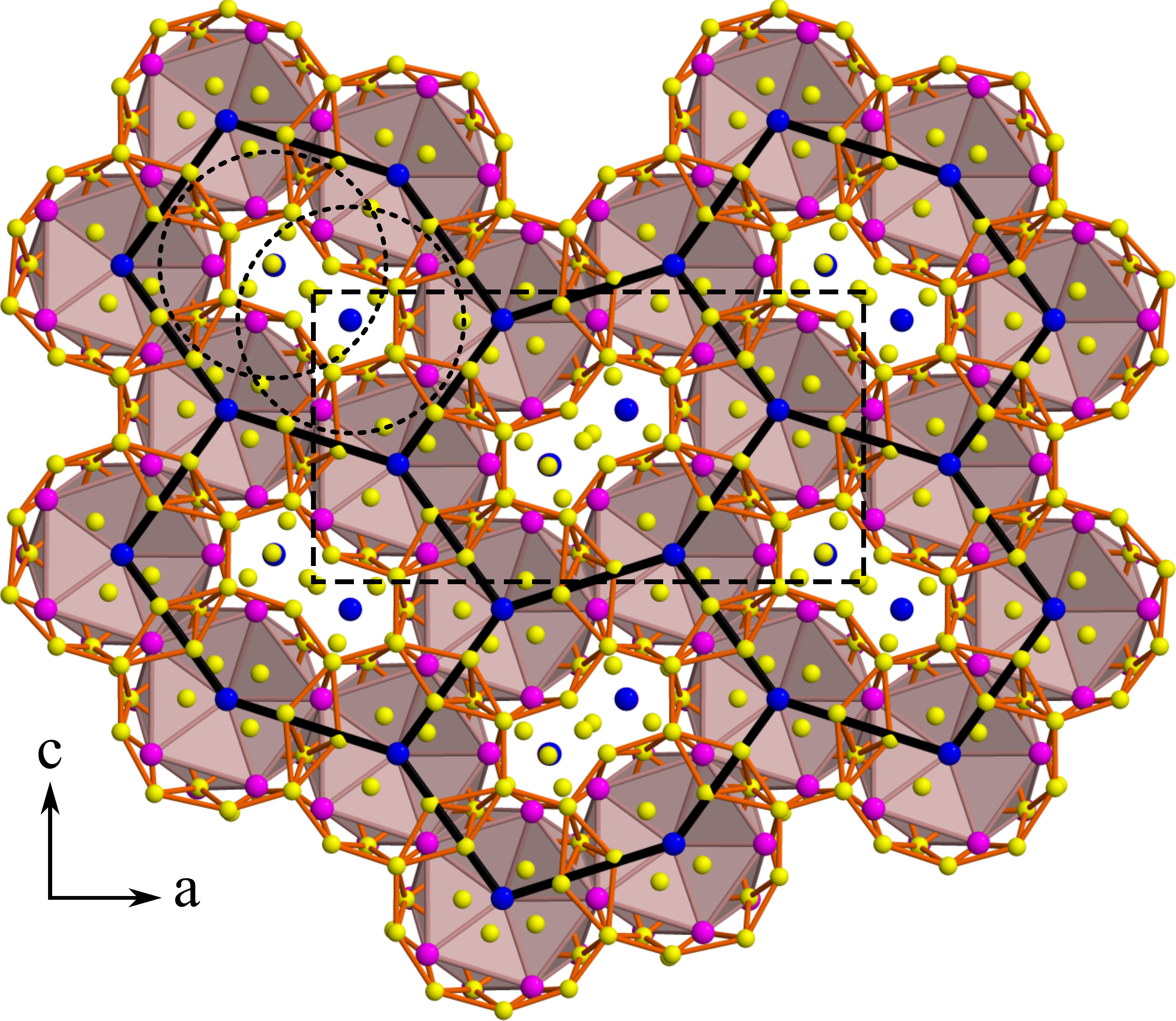}\\(a)\\

  \vspace{0.3cm}
  \centering
  \includegraphics[width=0.8\linewidth]{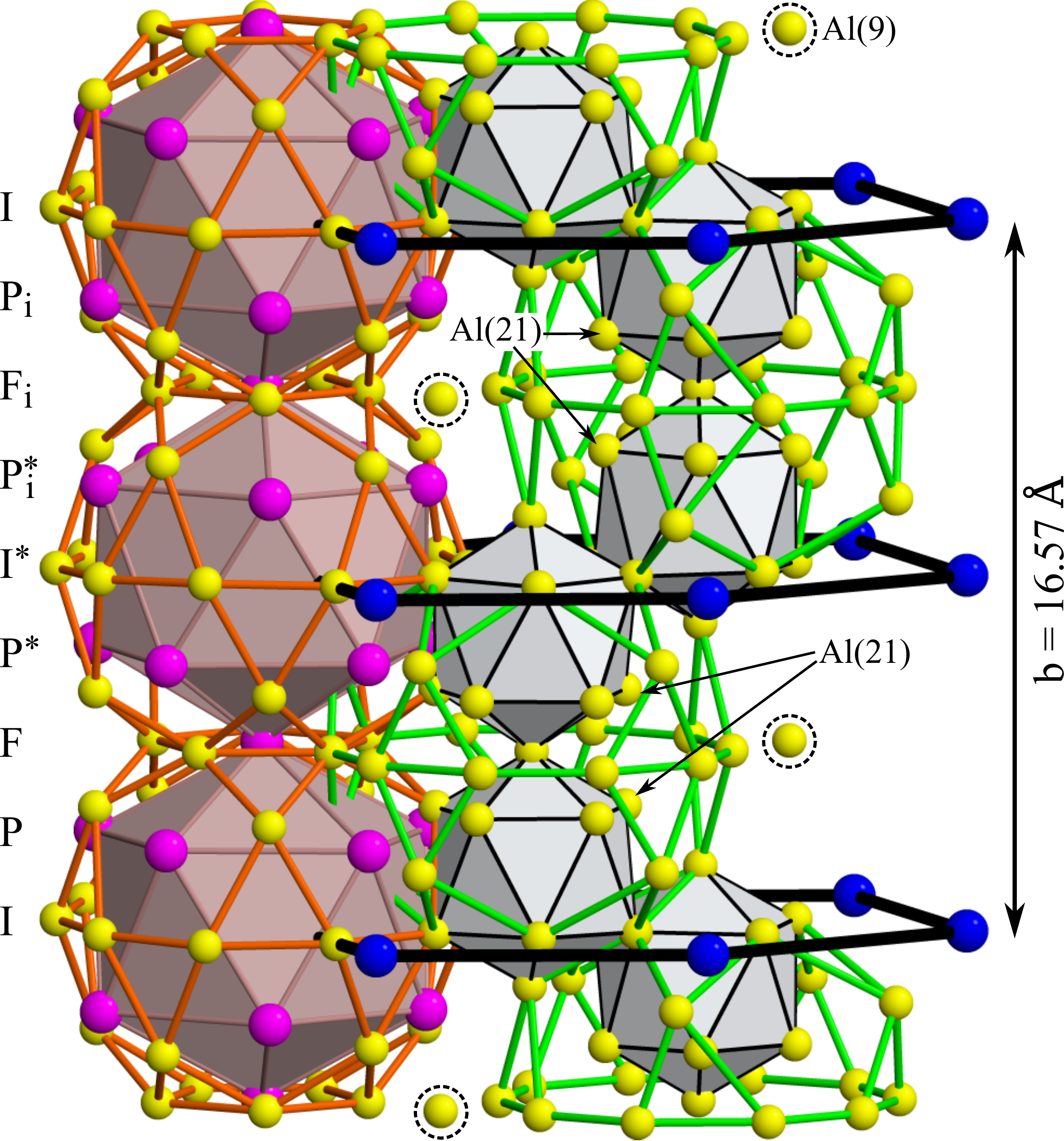}\\(b)
  \caption{(Color online) Alignment of the atomic clusters in $\xi'$. (a)
    Projection of the structure along the stacking axis of the PMI
    clusters. Black lines connect adjacent PMI-cluster columns. Dashed
    rectangle outlines the unit cell, dotted circles indicate the positions of
    the LBPP clusters. (b) Side view of one flattened hexagon. For the sake of
    simplicity, only one PMI column is shown. The Pd$_{10}$ pentagonal prisms
    are omitted from the LBPP clusters in this picture. The ``free'' Al atoms
    not covered by any cluster are encircled by black-dashed lines.}
  \label{fig:boudard_xipr}
\end{figure}

The PMI clusters are stacked on top of each other forming cluster columns
along the {\bf b} direction. Connecting adjacent columns results in a
two-dimensional tiling consisting of flattened hexagons with an edge length of
about 7.6~\AA. Fig.~\ref{fig:boudard_xipr}(a) shows a cut through the
structure perpendicular to the stacking axis. The LBPP clusters are located
inside the hexagons and are arranged in a zig-zag pattern along the {\bf b}
direction as shown in Fig.~\ref{fig:boudard_xipr}(b).

This description comprises more than 98\% of the atoms in the structure. Only
two Al atoms per hexagon (per unit cell height) cannot be attributed to any
cluster. 

The structure can be divided into three fundamental layers perpendicular to
the stacking axis: a flat (F) layer which is a mirror plane, a puckered (P)
layer, and an inverse (I) layer which contains the inversion center. The
remaining layers in Fig.~\ref{fig:boudard_xipr}(b) are obtained through
symmetry operations of the space group. The subscript i denotes inverse images
with respect to the I-layer, and the asterisk symbolizes mirror images with
respect to the F-layer. 

\section{Methods}
\label{sec:comp-methods}

\subsection{Tiling-Decoration Model}
\label{sec:tiling-decoration}
Our goal was to replace the average structure determined from X-ray
diffraction with many fractionally occupied or mixed sites by an ensemble of
configurations, each of which is a valid configuration with plausibly low
energy. 
Alternatively, recognizing the hierarchical organization of the average
structure in clusters, that fill the space following precisely obeyed linking
rules, our initial starting models are constructed by a tiling-decoration
method. The method creates a unique one-to-one correspondence between tilings
and atomic structures.\cite{mihalkovic_1996} This is achieved by the
decoration step: each tile, or more precisely each predefined tiling
object (node, linkage, tile interior...) carries an identical atomic motif
associated with the considered structure. Consequently, we can refine
positional parameters that are associated with each decorated tiling objects  
in a manner analogical to the refinement of Wyckoff positions in a unit cell
of a crystal. The long-range order of a tiling-decorated structure is then
defined by the underlying tiling.

\begin{figure}[h]
  \centering
  \begin{minipage}[b]{0.49\linewidth}
    \centering
    \includegraphics[scale=0.4]{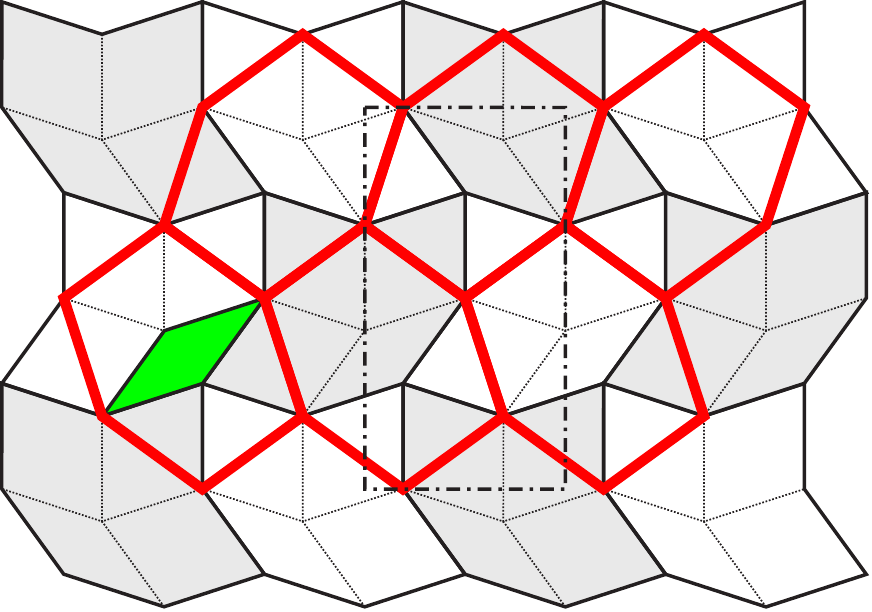}\\
    \centering $\xi'$
  \end{minipage}
  \begin{minipage}[b]{0.49\linewidth}
    \centering
    \includegraphics[scale=0.4]{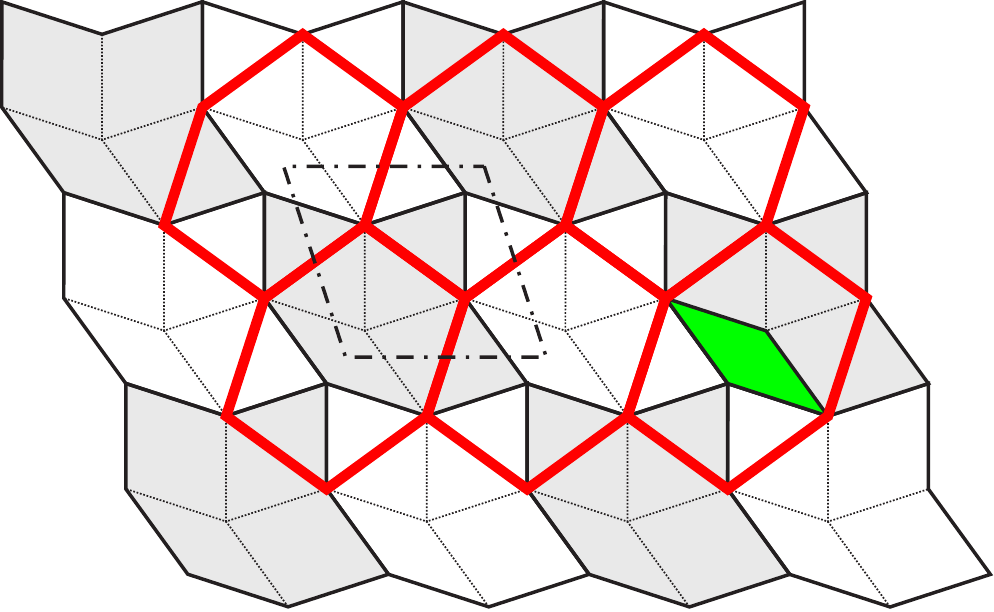}\\
    \centering $\xi$
  \end{minipage}

  \vspace{0.2cm}
  \centering
  \begin{minipage}[c]{0.32\linewidth}
    \centering
    \includegraphics[scale=0.45]{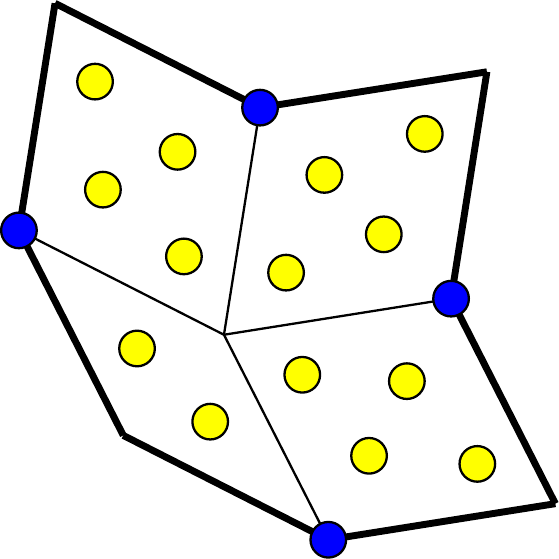}
  \end{minipage}
  \begin{minipage}[c]{0.32\linewidth}
    \centering
    \includegraphics[scale=0.45]{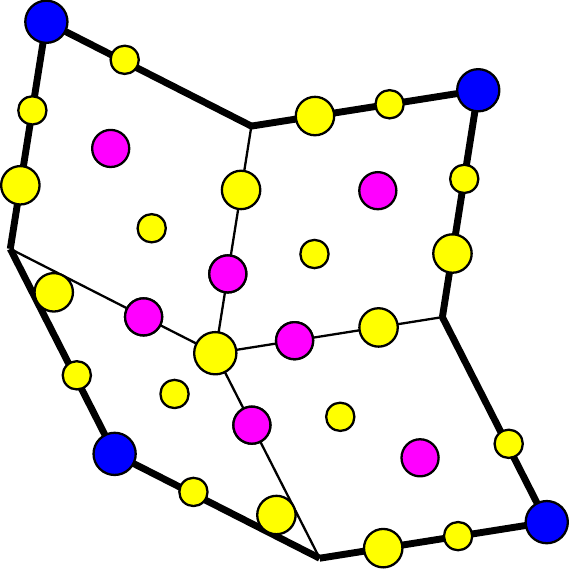}
  \end{minipage}
  \begin{minipage}[c]{0.32\linewidth}
    \centering
    \includegraphics[scale=0.45]{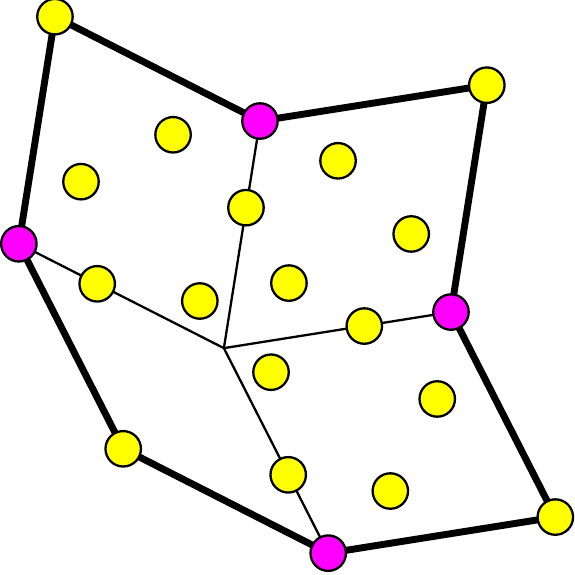}
  \end{minipage}
  \centering
  \begin{minipage}[b]{0.32\linewidth}
    \centering I-Layer
  \end{minipage}
  \begin{minipage}[b]{0.32\linewidth}
    \centering P-Layer  
  \end{minipage}
  \begin{minipage}[b]{0.32\linewidth}
    \centering F-Layer
  \end{minipage}
  \caption{(Color online) Top: HBS tiling models for $\xi$ and $\xi'$. Red
    (thick) lines
    connect adjacent PMI-cluster columns. Unit cells are highlighted with
    dash-dotted lines. Bottom: Tiling-decoration rule for the Boat tiles
    reproducing the Boudard model. Larger circles indicate atoms sitting in
    lower layers.}
  \label{fig:hbstilings}
\end{figure}

One benefit of this approach is a straightforward use of the $\xi$-phase as a
replacement of the twice as big $\xi'$-phase in the refinement process: we
simply refine the decoration motifs in $\xi$ by total energy calculations, and
afterwards substitute the underlying tiling representing $\xi$ by a tiling
representing $\xi'$. The tilings for $\xi$ and $\xi'$ are shown in
Fig.~\ref{fig:hbstilings}. In both cases the fundamental tiles are Boat tiles
(discussed below), packed either in parallel ($\xi$) or anti-parallel ($\xi'$)
order. 

{\it Decoration rule for the tiling.} We associate columns of PMI clusters
with the ``large'' $nodes$ of the binary Penrose tiling
(BPT).\cite{lancon_1988} The zig-zag columns of the LBPP clusters project 
exactly as ``small'' vertices of the BPT. We find that the atomic motifs are
in fact consistent with a particular subset of rhombic BPT family, namely the
so-called Hexagon-Boat-Star (HBS) tiling. As shown in
Fig.~\ref{fig:hbstilings}, the $\xi$- and $\xi'$-phases are pure Boat tilings
with no Hexagon or Star tiles. The Penrose rhombi, into which the H, B and S
tiles can be uniquely decomposed, have an edge length of about 6.6~\AA. We
chose a stacking period of 16.5~\AA\ for all structures. The initial
decoration rule for the Boat tiles in the $\xi$- and $\xi'$-tiling was set to
closely reproduce the atomic structure found by Boudard {\it et al.} We
decorate the tiles for each atomic layer separately. It only is necessary to
set up decoration rules for the three fundamental layers I, P, and F, and
align them along the stacking axis (I at 0.0, P at 0.125 and F at 0.25, in
units of the unit cell height). The decoration rules for the other layers are
obtained through symmetry operations applied to columnar atomic motifs bound
to our set of tiling objects. Our decoration rules relating the
Boudard model of $\xi'$ to the Boat tiles are shown in
Fig.~\ref{fig:hbstilings}.

A generalization of these pure-Boat tiling models to the family of HBS tilings
requires us to optimize one more decoration motif, occuring in the Hexagon tile,
when a pair of skinny rhombi shares a common edge. In our case such a
situation leads to a new type of horizontal interaction between the LBPP
clusters, and must therefore be optimized separately. This final decoration
rule was optimized (see Sec.~\ref{sec:phason-lines}), and we believe it
describes faithfully the atomic structures corresponding to the various
experimentally observed tiling patterns.\cite{heggen_structural_2008}

\subsection{Total Energy Calculations}
\label{sec:tot-energy-calc}
The total energy of a structure is obtained using a combination of classical
molecular dynamics and density functional theory (DFT). First, the candidate
structure is relaxed into its global energy minimum in a molecular dynamics
annealing approach. Typically, the sample is heated up to 1000~K and slowly
cooled down to zero Kelvin. We found that an annealing time of about 50~ps is
sufficient for our purposes. The annealing simulations are performed with the
ITAP molecular dynamics program {\sc IMD} in the NVT
ensemble.\cite{stadler_1997} Atomic interactions are modeled with the embedded
atom method (EAM).\cite{daw_1983} Suitable potentials have been developed for
this work with the force-matching method~\cite{ercolessi_1994} using the {\sc
  potfit} program.\cite{brommer_2007} A detailed account of the procedure can
be found in Ref.~\onlinecite{schopf_2010}.

The final total energies are calculated with the Vienna {\it ab initio}
simulation package {\sc VASP}.\cite{kresse_1993,kresse_1996} Our calculations
employ the projector-augmented wave (PAW) method~\cite{kresse_1999,paw_gga} in
the Perdew-Wang generalized gradient approximation (PW91-GGA).\cite{perdew_1992}
The Brillouin zone is sampled with a Monkhorst-Pack grid.\cite{monkhorst_1976}
The {\it k}-point density is chosen in such a way that all energies
converge to a precision of $10^{-3}$~eV. Using a conjugate gradient algorithm,
atomic positions as well as the shape and size of the unit cell are relaxed
once more, until an accuracy of $10^{-3}$~eV or better is
reached.\footnote{For the electronic minimization we use the blocked Davidson
  iteration scheme.} All calculations are performed with a constant
plane-wave energy cut-off of 269~eV.

To assess a low-temperature Al--Pd--Mn phase diagram, we calculate total
energies of all experimentally known stable phases including its binary and
pure elemental subsystems. We define the enthalpy of formation $\Delta H$ as
the difference of the total energy of a structure relative to the
composition-weighted total energies of its pure elements. Using the program
{\sc Qhull},\cite{Barber_1996} a convex hull of enthalpy versus composition
is calculated. Structures that minimize the enthalpy are considered
thermodynamically stable (at $T$=0~K). These phases constitute the vertices of
the convex hull. Edges and facets indicate two-phase and three-phase regions,
respectively. Structures with $\Delta H$ above this "enthalpy surface" are
unstable. The energy difference $\Delta E$ to the convex hull indicates
their degree of instability. More details of similar {\sc VASP} and convex
hull calculations can be found, for instance, in
Ref.~\onlinecite{mihalkovic_2004,mihalkovic_2007}.

When comparing unrelated structures, our target accuracy for the total
energies is 1~meV/atom, and this accuracy is certainly guaranteed by our
convergence criteria.  In our experience, using for example the
Perdew-Burke-Ernzerhof generalized gradient approximation
(PBE-GGA),\cite{PBE-GGA_1996} the energy differences $\Delta E$ from PW91-GGA
are usually reproducible to within 1--3~meV/atom. 
However, when it comes to comparing energy differences for variants of the same
structure, the energy differences should not, in general, be divided
by the number of atoms, since they only reflect differences in particular
places of the structure. Such energy differences can be reliably evaluated
with much higher accuracy than 1--3~meV/atom.

\newpage

\section{Results}
\label{sec:results}
\subsection{Low-Temperature Al--Pd--Mn Phase Diagram}
\label{sec:alpdmn-phase-diagram}
\begin{figure}[h]
  \centering
  \fbox{\includegraphics[width=0.95\linewidth]{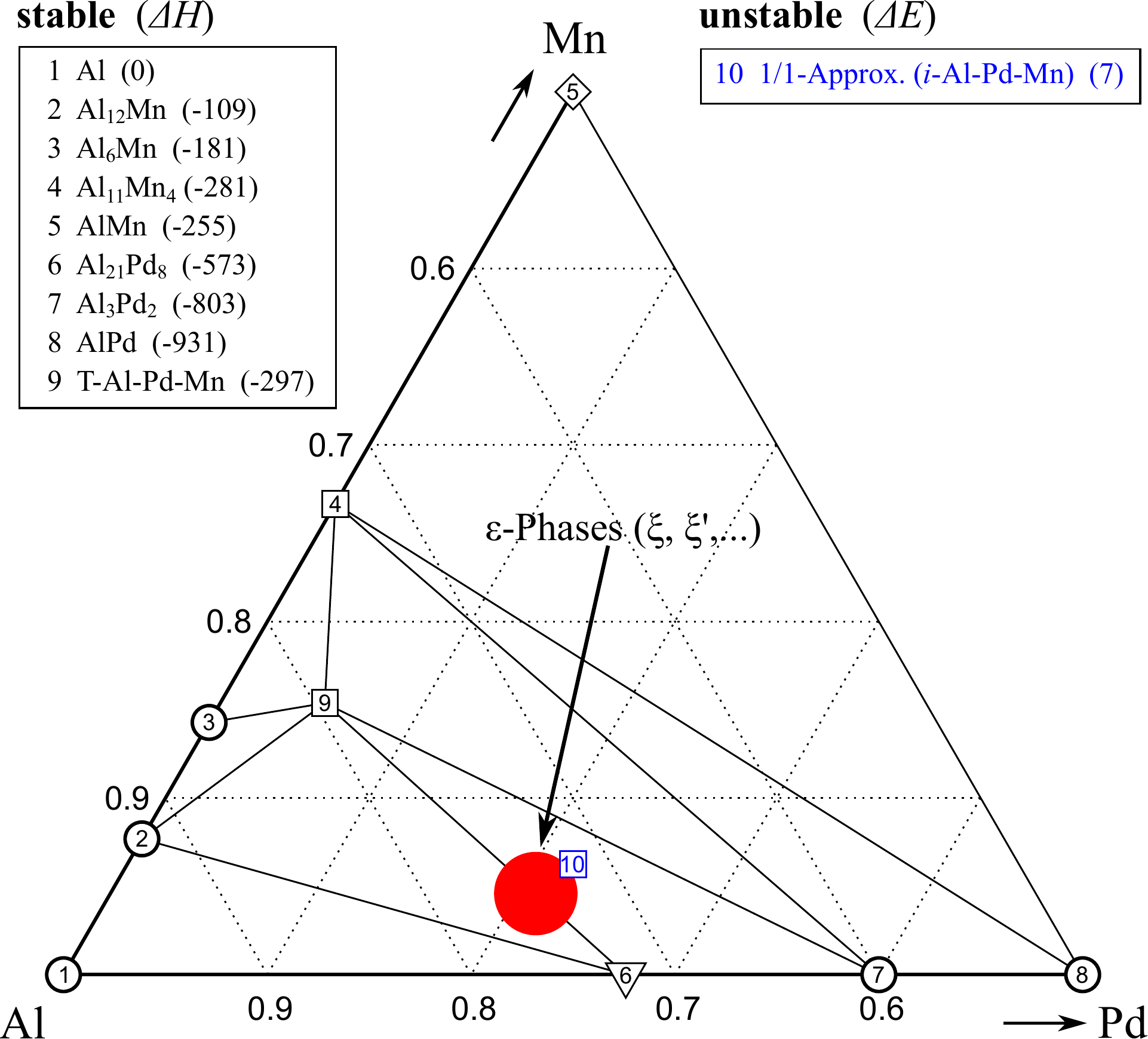}}
  \caption{(Color online) The Al--Pd--Mn energy-phase diagram in the Al-rich
    corner at $T$=0~K. Circles label known stable binary phases, diamonds
    indicate known metastable phases, squares are either unreported or unknown
    structures, and triangles correspond to high-pressure phases. Tie-lines
    connect low-enthalpy phases, constituting the convex hull vertices. 
    Enthalpies $\Delta H$ and energy differences $\Delta E$ to the convex hull
    are given in parantheses in meV/atom. The red spot outlines the approximate 
    compositions of the $\varepsilon$-phases.}
  \label{fig:alpdmn_phase_diagram}
\end{figure}

The binary Al-rich Al--Mn and Al--Pd phases are the main constraints for the
formation of Al-rich ternary phases. 
In the following, we first discuss the binary subsystems as well as all
previously known stable ternary phases. These phases are summarized in
Tab.~\ref{tab:competing-phases} along with the elementary subsystems. 
Fig.~\ref{fig:alpdmn_phase_diagram} shows the zero-temperature ``energy-phase
diagram'' in the Al-rich corner without our optimized $\xi'$-phases. 

{\em Al--Pd binary phases.} 
The stable Al-rich phases are Al$_4$Pd, Al$_{21}$Pd$_8$, Al$_3$Pd$_2$, and
AlPd.\cite{Pearson_handbook} The latter is a FeSi-type structure. The usual
$B2$-type phase (CsCl) with the same composition becomes stable only at high
temperature. Al$_3$Pd$_2$ is an ordered-vacancy state related to $B2$-AlPd,
where each Pd atom has only three Pd nearest neighbors. The Al$_{21}$Pd$_8$
phase has a more complex structure, and an Al-content similar to the
$\xi'$-phase. One efficient structural diagnostics for Al-rich Al--TM
(transition metal) compounds is to look at the local environments of the TM
atom. These structures often seek to maximize the number of Al--TM bonds. TM
atoms tend to have a few or no TM nearest neighbors. Looking at the
environments of all TM atoms, (almost) all atoms of the structure are
represented. Both Pd sites in Al$_{21}$Pd$_8$ have a very similar local
environment of 10~Al atoms. These local environments are reminiscent of the
inner Al shells in the PMI clusters of the $\xi'$-phase.
\begin{table*}
\caption{Low-temperature phases in the Al-rich corner of the Al--Pd--Mn phase
  diagram (cf.~Fig.~\ref{fig:alpdmn_phase_diagram}).} 
\begin{ruledtabular}
\begin{tabular}{rlccd}
& & & & \multicolumn{1}{c}{\textrm{Total Energy}} \\
\# & Structure & Space Group (No.) & Pearson Symbol &
\multicolumn{1}{c}{\textrm(eV/atom)}\\
\hline
1 & Al                & $Fm3m$ (225)      & $cF4$   & -3.688 \\
--& Pd                & $Fm3m$ (225)      & $cF4$   & -5.199 \\
--& Mn                & $I\bar{4}3m$ (217)& $cI58$  & -8.963 \\
2 & Al$_{12}$Mn       & $Im\bar{3}$ (204) & $cI26$  & -4.202 \\
3 & Al$_{6}$Mn        & $Cmcm$ (63)       & $oC28$  & -4.623 \\
4 & Al$_{11}$Mn$_{4}$ & $P\bar{1}$ (2)    & $aP15$  & -5.372 \\
5 & AlMn              & $P4/mmm$ (123)    & $tP4$   & -6.581 \\
6 & Al$_{21}$Pd$_{8}$ & $I4_{1}/a$ (88)   & $tI116$ & -4.674 \\
7 & Al$_{3}$Pd$_{2}$  & $P\bar{3}m1$ (164)& $hP5$   & -5.091 \\
8 & AlPd              & $P2_{1}3$ (198)   & $cP8$   & -5.369 \\
9 & T-Al$_{31}$Pd$_{2}$Mn$_{6}$ & $Pnma$ (62) & $oP156$ & -4.873 \\
10& 1/1-$i$-Al$_{23}$Pd$_{7}$Mn$_{2}$ & $P2_{1}2_{1}2_{1}$ (19) & $oP128$ &
-4.863\\
\end{tabular}
\end{ruledtabular}
\label{tab:competing-phases}
\end{table*}

While the $\lambda$-Al$_{4}$Pd phase is reported as a stable Al--Pd
compound,\cite{okamoto_2003} its structure has never been determined. However,
the recently determined structure of the Al$_4$Pt was conjectured to be
isostructural with it.\cite{Woerle_2008} Substituting Pt for Pd in this
hexagonal structure, and resolving obvious correlations between four
fractionally occupied Al sites, we find two variants of this phase just
slightly unstable (by about 5~meV/atom) against decomposition to $fcc$ Al and
Al$_{21}$Pd$_8$. The structure of Al$_4$Pd is similar to the
Al$_{21}$Pd$_8$: both are packings of similar Al$_{10}$Pd clusters.

At high temperature, the $\xi'$-phase extends into the Al--Pd binary
system.\cite{matsuo_1994,Grushko_1999,Klein_2000} In Sec.~\ref{sec:al3pd} an
optimized low-temperature structure model for Al$_{3}$Pd is presented and
compared with the $\xi'$-phase.

{\em Al--Mn binary phases.} The evaluation of phase stabilities in the Al--Mn
binary phase diagram (based on {\it ab initio} total energy calculations)
results in two sub-optimally resolved phases: (i) Al$_8$Mn$_5$ phase: it is
difficult to refine simultaneously the Al--Mn chemical ordering along with a
magnetic structure; (ii) $\mu$-Al$_4$Mn contains approximately 570 atoms in a
hexagonal unit cell. Due to the reported mixed/partial occupancies, an
optimization of the low-temperature structure presents a significant challenge
to the available computational resources. Thus, our calculated Al--Mn convex
hull includes four phases: Al$_{12}$Mn (Al$_{12}$W prototype), Al$_{6}$Mn,
Al$_{11}$Mn$_{4}$, and tetragonal AlMn.\cite{alloydata} Out of these, the
lowest formation enthalpy $\Delta H$ is achieved by the Al$_{11}$Mn$_{4}$
phase, while Al$_{6}$Mn and Al$_{12}$Mn enthalpies are barely below the
tie-line connecting Al$_{11}$Mn$_{4}$ with $fcc$ Al.


Since the $\xi'$-phase (and all other $\varepsilon$-phases) is located closer
to the Al--Pd binary system in the phase diagram, we believe that the Al--Mn
binary system has a smaller impact on these phases. Even though the Al$_4$Mn
and Al$_8$Mn$_5$ phases are refined only suboptimally, the lowest-enthalpy
structure is Al$_{11}$Mn$_4$. It has also a more similar Al content to $\xi'$,
and should therefore be the most important Al--Mn phase to consider.

{\em Previously known stable ternary phases} include the
icosahedral\cite{tsai_1990,boudard_1992} and the decagonal\cite{beeli_1991}
quasicrystalline phases, the $\xi'$-phase, and the so-called T-phase, an
approximant of the decagonal phase with 12~\AA\ stacking 
period.\cite{klein_1997, beraha_1998} The $\xi'$-phase has a composition
similar to the icosahedral phase, and has been treated as its
approximant.\cite{beraha_correlated_1997} At the same time it is a proper 
approximant of the decagonal phase with 16~\AA\ periodicity. 

The stable icosahedral quasicrystalline phase has a composition of
Al$_{71.1}$Pd$_{20.2}$Mn$_{8.7}$.\cite{gratias_2002} Here we represent this
phase by the best (lowest energy) approximant model available to us, containing
128 atoms in a cubic unit cell with an edge length of about 12.6~\AA, and
a composition of Al$_{71.9}$Pd$_{21.9}$Mn$_{6.3}$.\cite{mihalkovic-unpubl} The
structure model is conventionally denoted as ``1/1''. All atoms are
incorporated in the four PMI clusters and the four Al$_{12}$Pd icosahedra,
decorating the vertices of a canonical cell tiling.\cite{henley_1991} The
innermost shells of the PMI clusters in this icosahedral approximant are
comprised of 10 Al atoms, as opposed to our optimal $\xi'$-phases, that
contain only 9 Al atoms in each inner PMI shell. In our final evaluation of
the phase stability, this 1/1 approximant is unstable by only 7 meV/atom. As a
competing model, we also computed total energies of two models reported in
literature, but both had significantly higher
energies.\cite{quandt_2000,*elser_1996, Zijlstra_2005, *Zijlstra_2005-2}

Currently, there is no good estimate of total energies for the 12~\AA\
decagonal phase. In our database, the structure is represented by its
approximant T-Al--Pd--Mn. In agreement with recent experimental
work\cite{Balanetskyy_2008} composition and chemical ordering of the optimal
low-temperature ternary structure is
Al$_{31}$Pd$_{2}$Mn$_{6}$.\cite{mihalkovic-unpubl2} Both space group and the
number of atoms in the unit cell are identical with the reported binary
Al$_{3}$Mn high-temperature phase.\cite{hiraga_1993}

The approximate compositions of the $\varepsilon$-phases are shown in
Fig.~\ref{fig:alpdmn_phase_diagram}. Thus, our competing phases are
T-Al--Pd--Mn, Al$_{21}$Pd$_{8}$, and either Al$_{12}$Mn or Al$_{3}$Pd$_{2}$. 
Depending on the composition of the candidate structure, we will refer the
energy difference $\Delta E$ to either tie-triangle. 
Fig.~\ref{fig:phase-diagram-new} includes the most important low-energy
structures we optimized in this work. All optimized structures are further
summarized in Tab.~\ref{tab:low-temp-xiphases}.

In addition to the reported ternary phases, we also include a ternary version
of the recently discovered quaternary structure 
Al$_{72}$Pd$_{18}$Mn$_5$Si$_5$,\cite{simura_2011} with Si substituted by Al. 
According to our stability evaluations at $T$=0~K, the structure is
insignificantly more stable than our optimized $\xi$- and $\xi'$-phases, as
shown in Fig.~\ref{fig:phase-diagram-new}, and is discussed at the end of
Sec.~\ref{sec:discussion}.

\subsection{Structural Optimization of \texorpdfstring{$\xi$}{xi} and
 \texorpdfstring{$\xi'$}{xi'} with respect to \texorpdfstring{$T$}{T}=0~K total
energies}
\label{sec:Optimization}
Our structure refinement is divided into two steps. In a first step we
analyze the occupancies and occupancy correlations within the inner PMI shells.
Determining the correct number of atoms contained within the inner shells is
an important task, in which intuition easily fails. It is well known that
quasicrystal-related Al--TM compounds often exhibit deep pseudogaps. The
position of the Fermi level relative to the pseudogap center may have an
important impact on the total energy (as we indeed show below), and is not
dictated necessarily by local constraints. At the same time, the occupancy
factors refined from diffraction data cannot be taken for granted, due to the
couplings between Debye-Waller and occupancy factors. The loose inner shells
with low symmetry encapsulated within the highly symmetrical outer shells of
the PMI clusters may possess nearly continuous degrees of freedom connecting
rather closely spaced alternative orientations of this inner cluster. A mix of
such configurations is nearly impossible to represent by a few discrete sites
available to a tractable refinement.

In a second step, we vary the chemical ordering within the LBPP clusters. To
make the optimization tractable, we assumed that the variations of the LBPP
cluster especially on its 5-fold axis are in the first-order approximation
independent from the inner-PMI-shell cluster variations. As we show later,
careful optimization of the chemical ordering along this cluster axis revealed
striking impact on the total energy, and proved that the $\xi$- and
$\xi'$-phases are stable down to zero temperature, according to our DFT
approximated energetics. One of the counter-intuitive findings was that the
optimal chemical sequence along the LBPP 5-fold axis places Mn--Pd atoms
as nearest neighbors -- a feature that we have never seen in other low-energy
structures of the ternary Al--Pd--Mn compounds.

Finally, as already mentioned in Sec.~\ref{sec:boudard-model}, there are some
atomic sites reported in the diffraction refinement, located in the
intersticial spaces between the PMI and LBPP clusters. Placement of these
atoms is in apparent conflict with other Al atoms in the second PMI shells. We
could never find a plausible low-energy configuration that would occupy these
sites. 

\subsubsection{Occupancies of inner PMI shells}
\label{sec:occupancy-innermost}
In this section, we systematically minimize the total energy of the $\xi$-phase
with respect to (i) the number of Al atoms present in the inner PMI shells;
(ii) optimal shape/orientations of these inner subclusters. The inner shells
are encapsulated within nearly icosahedral second shells, and since they
comprise less than 12 atoms, their symmetry is necessarily lower than the
icosahedral one, leading to several possible orientations. The two constraints
selecting among these inner-shell orientations are vertical (along PMI
stacking axis) and horizontal (perpendicular to PMI stacking axis)
interactions between the PMI clusters. While the vertical interaction is
pretty much separable and independent from column to column, the horizontal
PMI--PMI interactions select the favourable spatial patterns of PMI columns.  

The monoclinic unit cell of $\xi$ contains four PMI clusters (two adjacent PMI
columns, each with two clusters on top of each other). We start occupying each
inner shell with 8~Al atoms, and gradually increase the number up to 11~Al
atoms per PMI. In the following the TM(8) and TM(11) sites of the LBPP
clusters are occupied by Mn and Al atoms, respectively, consistent with the
Boudard model [cf.~Fig.~\ref{fig:xi_clusters}(b)]. The resulting
structures contain 32~Pd and 8~Mn atoms, whereas the number of Al~atoms varies
from 110 for 4$\times$Al$_{8}$, up to 122 for an inner-shell occupation of
4$\times$Al$_{11}$. The relative stability of each model ($\Delta E$ from
the appropriate tie-triangle) is evaluated by constructing the convex hull from
the enthalpies of formation $\Delta H$ of all binary subsystems as well as all
ternary phases.

We found that the structures are most stable with either Al$_{9}$ or Al$_{10}$
shells within each PMI cluster. The models are unstable by $\Delta E = 13.9$ and
12.6~meV/atom, respectively. Slightly higher energies were obtained for
structures containing both Al$_{9}$ and Al$_{10}$ shells, indicating an
unfavourable interaction between both shells. The PMI clusters with Al$_{8}$
or Al$_{11}$ inner shells are highly unfavourable. These structures were
unstable by 42.5 and 33.6~meV/atom, respectively. The Al$_8$ inner shells 
were even unstable mechanically, since during the annealing process some atoms
from the outer Al$_{30}$ shells drifted inward, forming Al$_9$ shells.
On the other hand, the inner shells seemed to be ``over-occupied'' with 11~Al
atoms, as some atoms drifted toward the outer PMI shells in the course of the
molecular dynamics annealing.
\begin{figure}[h]
  \centering
  \begin{minipage}[b]{0.49\linewidth}
    \includegraphics[scale=0.12]{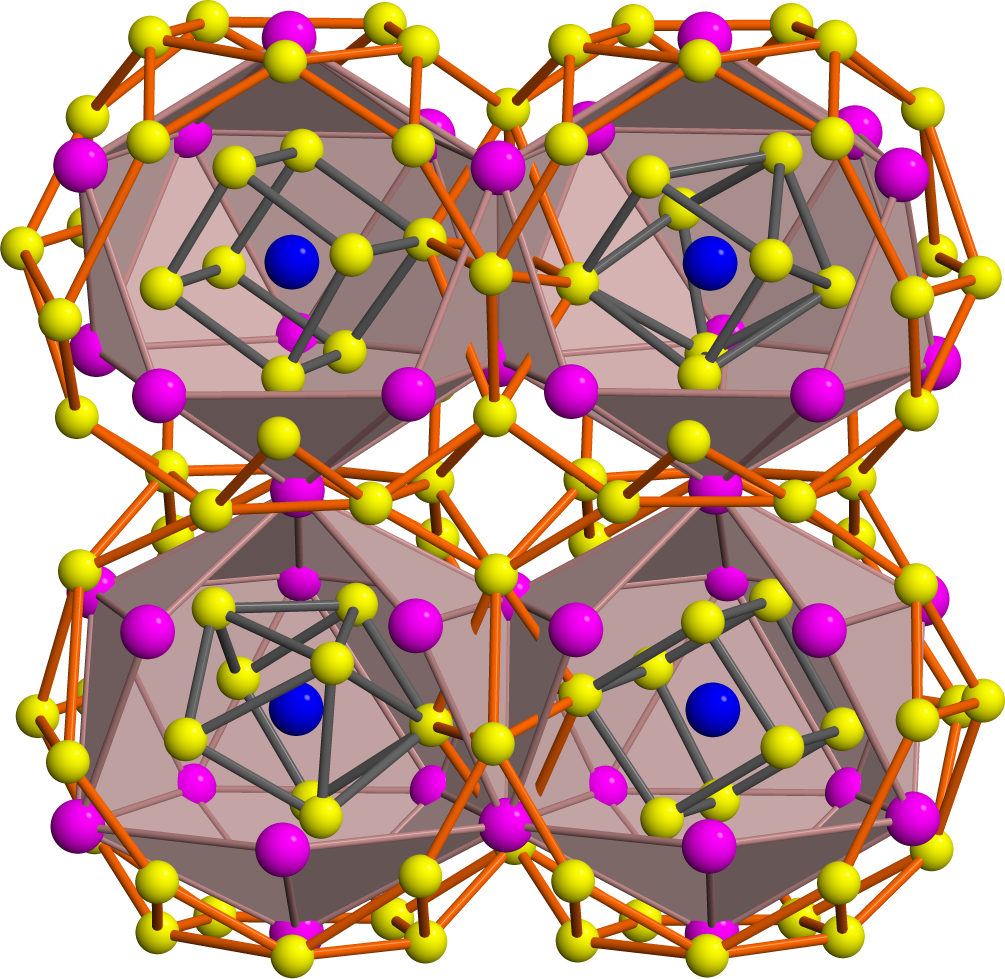}\\
    \centering 4$\times$Al$_{8}$
  \end{minipage}
  \begin{minipage}[b]{0.49\linewidth}
    \includegraphics[scale=0.12]{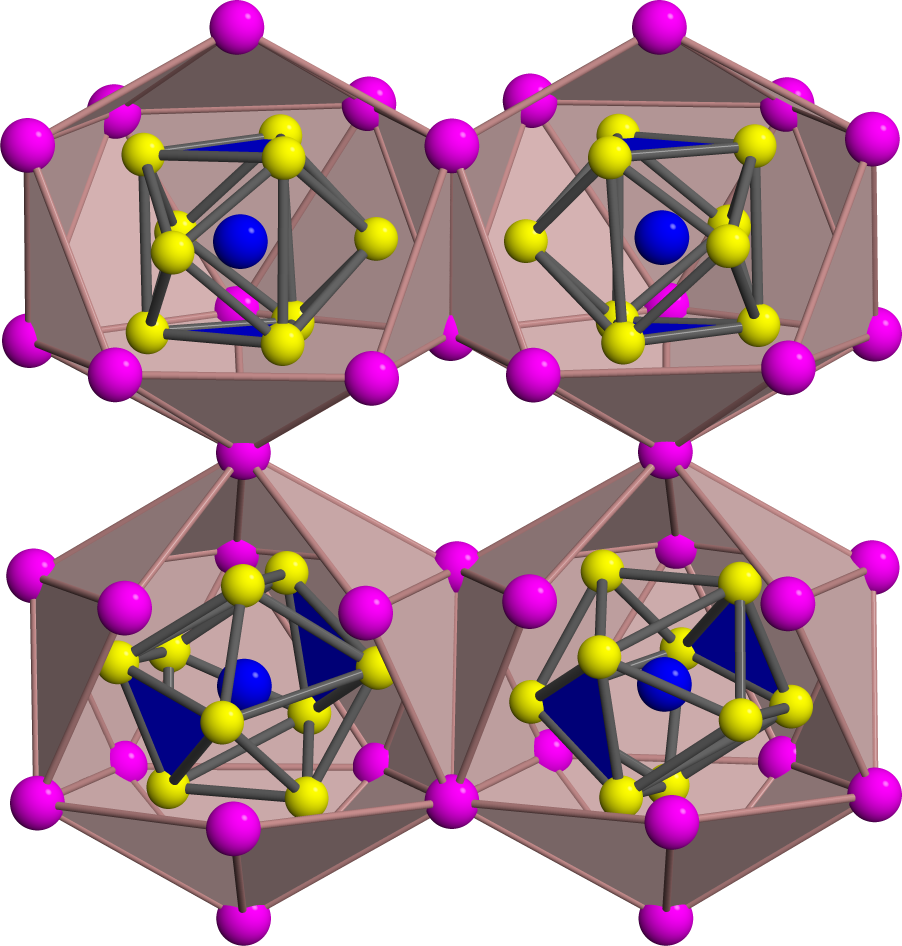}\\
    \centering 4$\times$Al$_{9}$
  \end{minipage}
  \centering
    \includegraphics[scale=0.12]{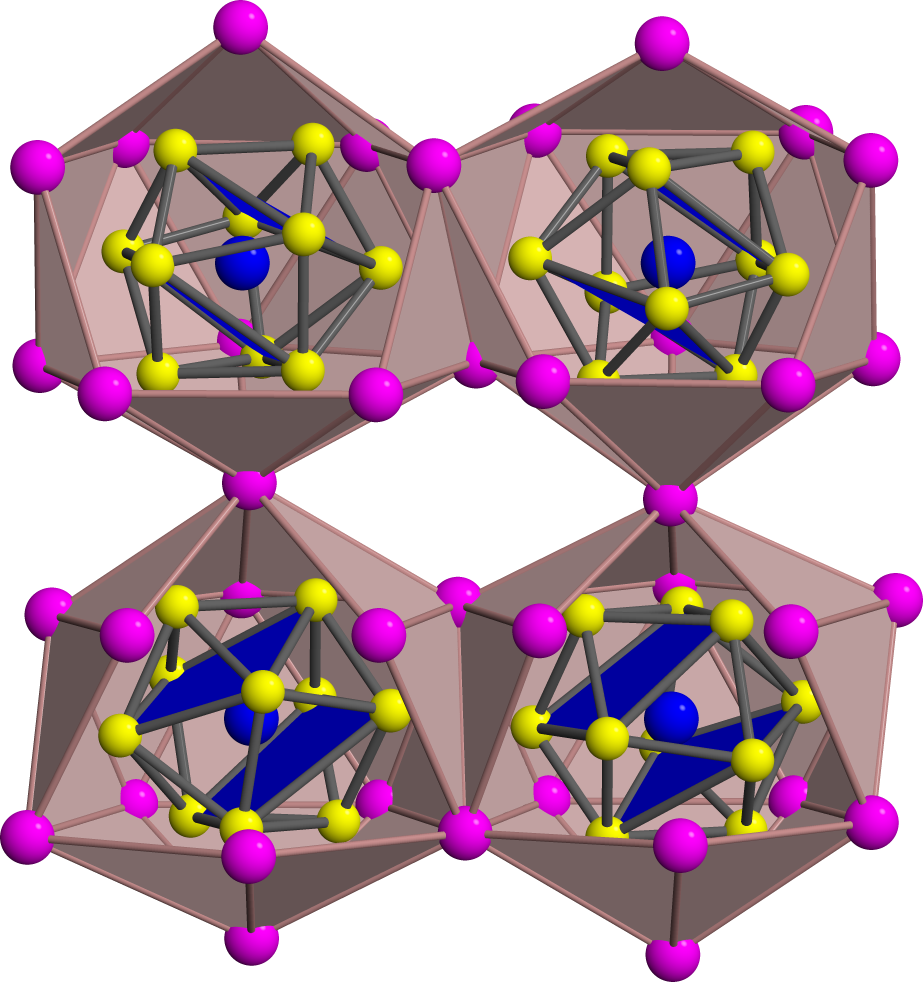}\\
    \centering 4$\times$Al$_{10}$
  \caption{(Color online) PMI clusters in $\xi$ after relaxation. The vertical
    axis corresponds to the PMI stacking axis. The outer Al$_{30}$ shells are
    shown only for the 4$\times$Al$_{8}$ occupation. The triangles of the
    trigonal Al$_{9}$ shell and the antiprisms of the Al$_{10}$ shells are
    shaded in blue to outline their 3-fold and 2-fold axes, respectively.}
  \label{fig:PMIs_innermost}
\end{figure}

{\it Shape of the inner shells.} The PMI clusters obtained after relaxation
are shown in Fig.~\ref{fig:PMIs_innermost}. We observed two different types of
Al$_{8}$ shells. One type of shell is a nearly perfect cube with edge lengths
of 2.7--3.0~\AA. The other type of shell can be described as a distorted
square antiprism. The Al$_{9}$ shell is a trigonal prism, capped within a
larger triangle (tricapped trigonal prism, TTP). Its 3-fold axis is aligned
(approximately) along the pseudo-decagonal (stacking) direction. The Al$_{10}$
shells are best described by bicapped square antiprisms (BSA), where the
2-fold axes are oriented along the 2-fold axes of the outer PMI shells. The
Al$_{9}$ and Al$_{10}$ shells can point in different directions as outlined in
Fig.~\ref{fig:PMIs_innermost}. For instance, in Fig.~\ref{fig:PMIs_innermost}
the 3-fold axis of the TTPs in the upper two PMI clusters point along the
stacking axis, whereas the lower TTPs are aligned along a local 5-fold
axis. For the Al$_{11}$ shells, no unique description can be established, as
the shells had variable shape and complicated pattern of orientational
relationship with respect to the second shell.

\subsubsection{Chemical ordering within LBPP clusters}
\label{sec:chem-LBPP}
The inner-shell optimization reported in the previous section was performed
using a fixed standard model for the chemical ordering within the LBPP
cluster. In this second stage of the refinement, our strategy is to fix the
energy-minimizing model for the inner shells, and refine the LBPP cluster
interior.  Due to the small energy difference between the structures
containing Al$_{9}$ and Al$_{10}$ shells, we accordingly split the LBPP
optimization into two branches. Since the Al$_9$ variant proved more optimal,
in the following, unless stated otherwise, the PMI clusters have Al$_9$ inner
shells.

\begin{figure}
  \centering
  \begin{minipage}[b]{0.32\linewidth}
    \centering
    \includegraphics[scale=0.14]{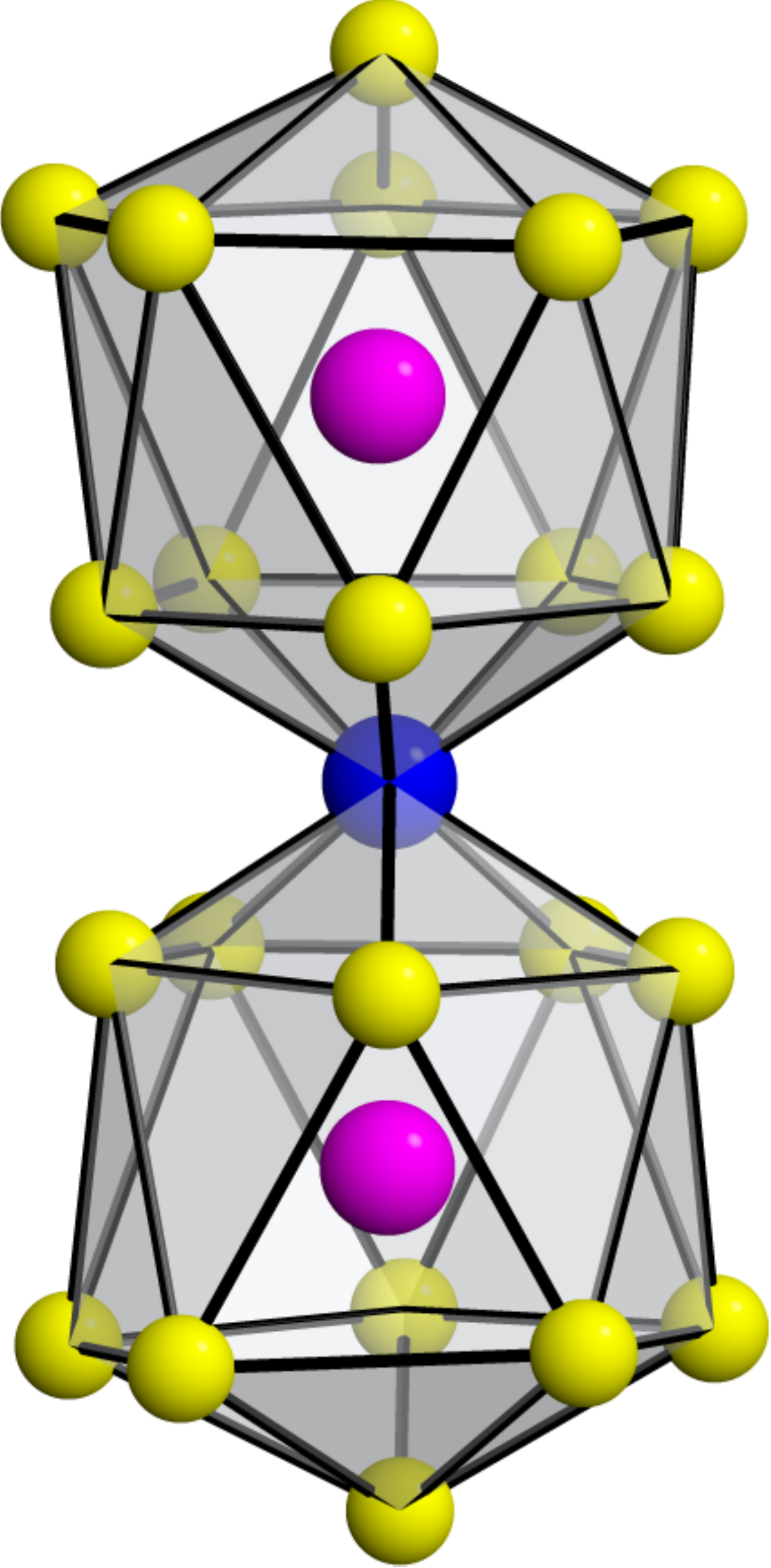}\\ (a)
  \end{minipage}
  \begin{minipage}[b]{0.32\linewidth}
    \centering
    \includegraphics[scale=0.14]{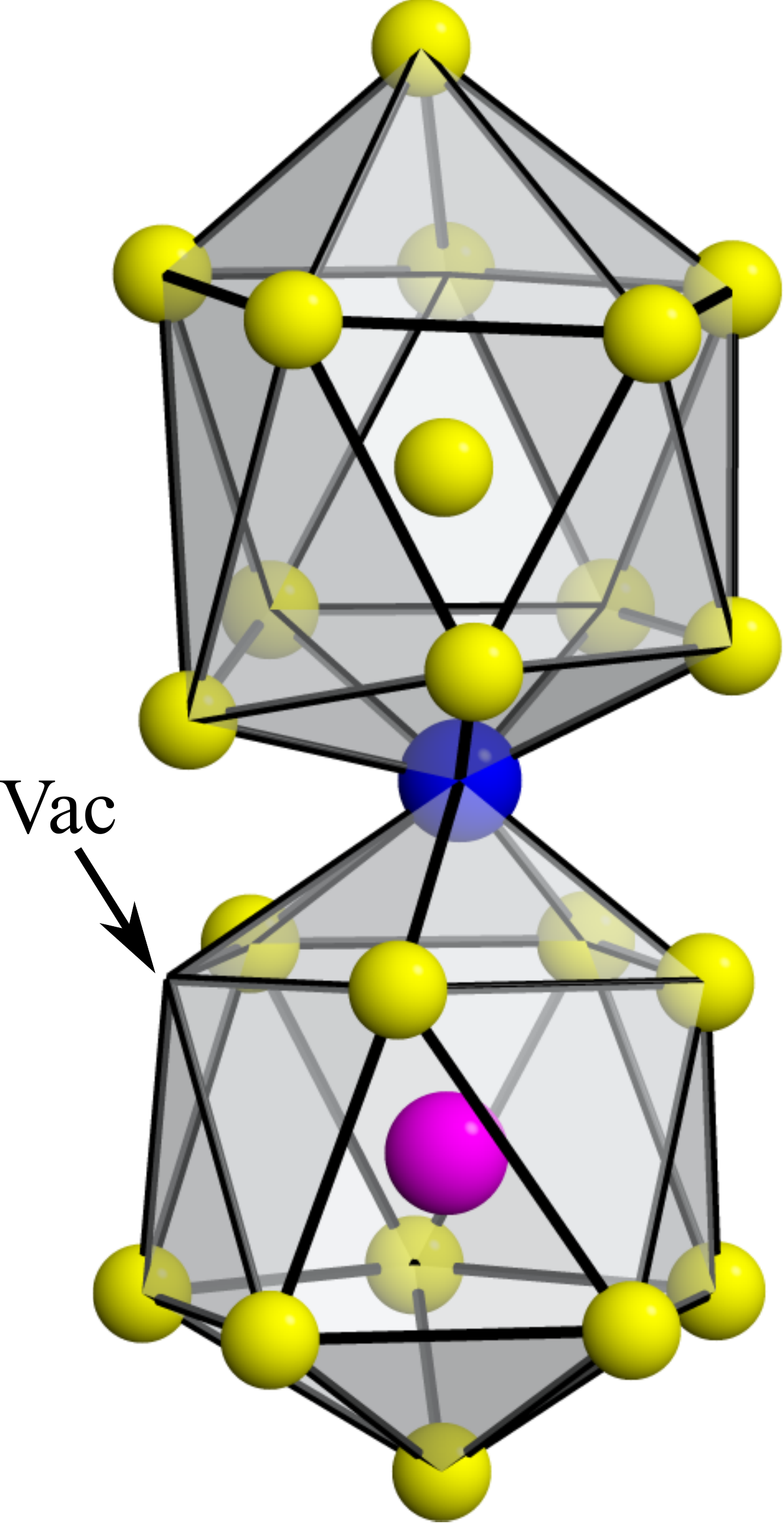}\\ (b)
  \end{minipage}
  \begin{minipage}[b]{0.32\linewidth}
    \centering
    \includegraphics[scale=0.14]{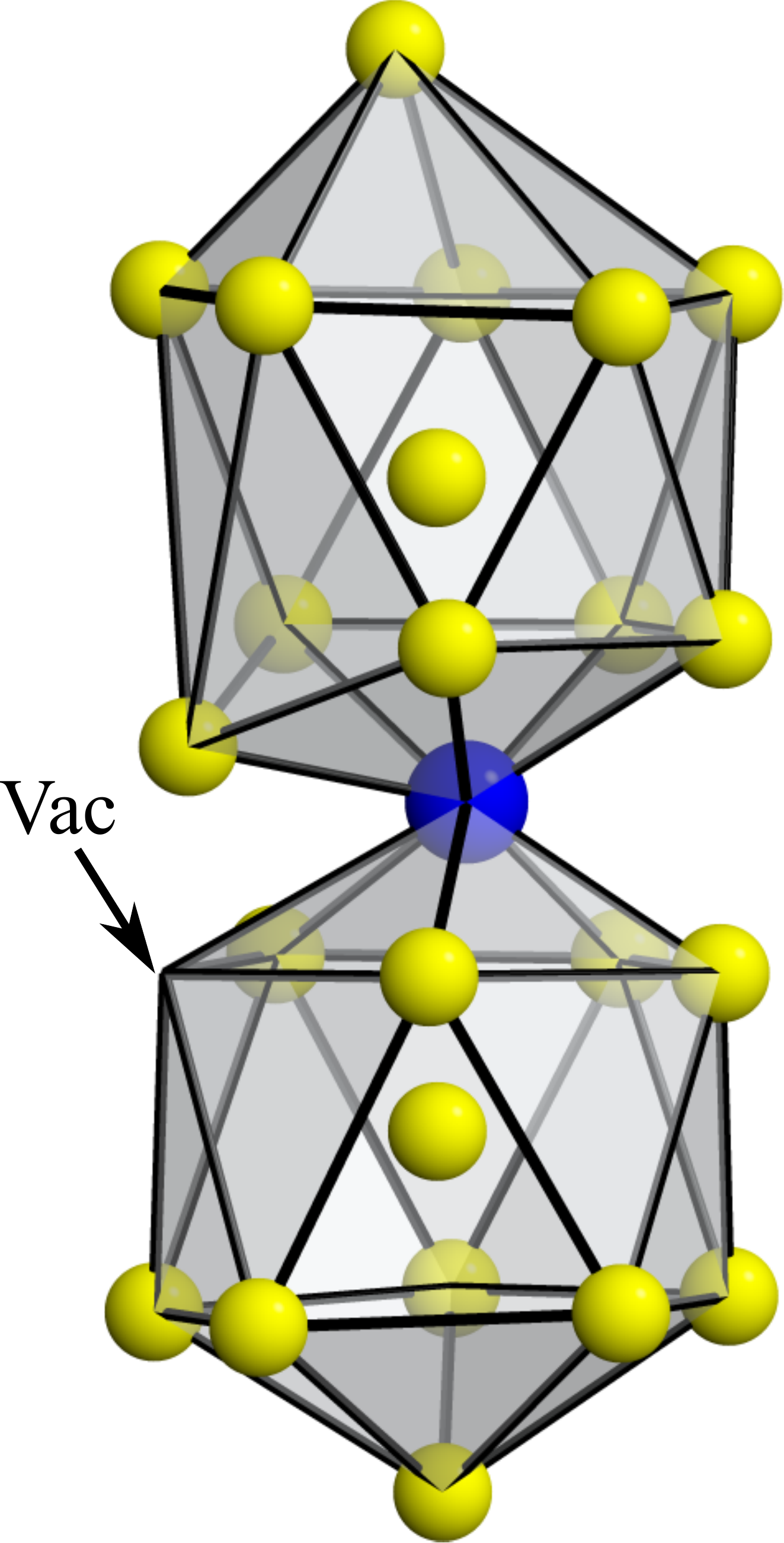}\\ (c)
  \end{minipage}

  \vspace{0.2cm}
  \includegraphics[width=1\linewidth]{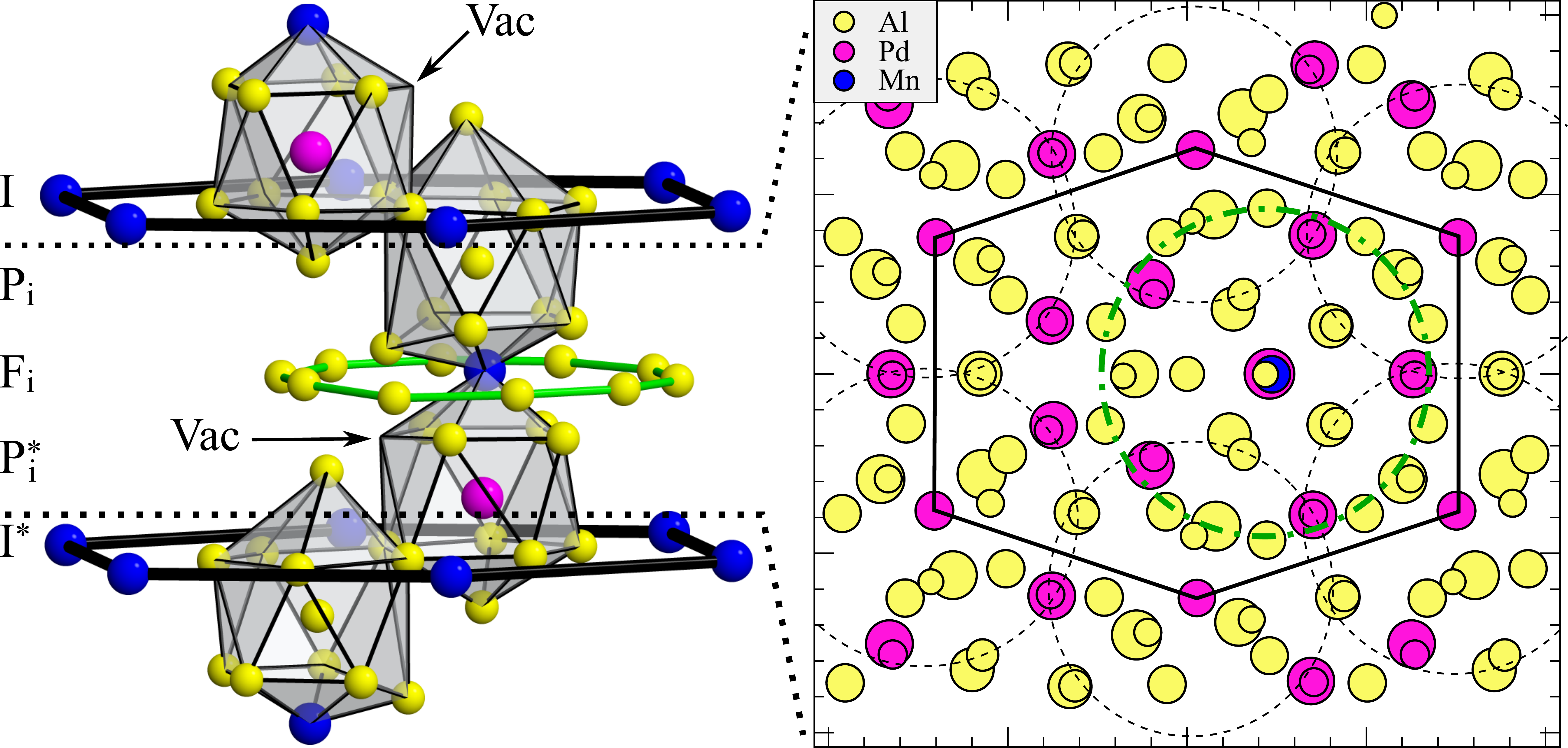}\\ (d)\\[0.2cm]
  \fbox{
  \includegraphics[scale=0.05]{Al.png}\hspace{0.2mm}Al \ \ 
  \includegraphics[scale=0.05]{Pd.png}\hspace{0.2mm}Pd \ \
  \includegraphics[scale=0.05]{Mn.png}\hspace{0.2mm}Mn}
  \caption{(Color online) (a)-(c) Icosahedra of the LBPP clusters with
    different chemical 
    orderings. The structure in (a) is unstable by 0.2~meV/atom. The
    structures in (b) and (c) are stable with respect to all competing
    phases. All three structures contain Al$_{9}$ inner shells in their PMI
    clusters. (d) Alignment of the LBPP cluster from (b)
    along the PMI stacking axis. Only atoms of the flat and adjacent puckered
    layers are  displayed in the right picture. Larger circles indicate atoms
    sitting in lower layers. PMI clusters are outlined with dashed
    circles. The Al decagon located on the flat layer is outlined with a
    dash-dotted circle. The unoccupied sites in (b) and (c) correspond to
    Al(21) in Ref.~\onlinecite{boudard_structure_1996}. Vac means vacancy.} 
  \label{fig:LBPPs}
\end{figure}

Our optimization first focuses on the atoms on the LBPP axis. As already
mentioned, the LBPP cluster can alternatively be described in terms of
separated spherical shells, similar to the PMI cluster. In both clusters the
second shell consists of two subshells comprised of Al and Pd atoms. The inner
part is either a pure Al$_{8-11}$ shell (PMI), or a composite Al$_{10}$Mn$_{2}$
shell (LBPP). We believe it is more appropriate to change the chemical
ordering within the LBPP cluster to obtain a similar atomic arrangement as in
the PMI cluster. Therefore, we occupy the central TM(11) position of the LBPP
cluster by Mn and replace the original Mn atoms located at the TM(8) sites by
Pd as shown in Fig.~\ref{fig:LBPPs}(a). In doing so, we found a structure
lying very close to the convex hull ($\Delta E = 0.2$~meV/atom). The
first-neighbor Mn--Pd distances within the LBPP clusters are 2.86~\AA\ and
seem to be the key factor for the enhanced stability. The space group of this
structure is $Cmcm$ (No.~63). The unit cell contains 154~atoms in the
primitive cell, with a composition Al$_{56}$Pd$_{18}$Mn$_{3}$.  For
comparison, the lowest-energy structure we found with the initial chemical
ordering [see.~Fig.~\ref{fig:xi_clusters}(b)] was unstable by $\Delta E =
12.6$~meV/atom.

We went one step further and broke the mirror symmetry of the LBPP cluster by
replacing one of the Pd atoms within the icosahedra by Al. During the annealing
certain Al atoms drifted toward the PMI clusters, resulting in some Al$_{10}$
inner shells, and leaving one atomic site in each LBPP clusters unoccupied as
shown in Fig.~\ref{fig:LBPPs}(b). The same behavior was observed when both
TM(8) sites were occupied by Al [see Fig.~\ref{fig:LBPPs}(c)]. The structures
were almost stable. We removed the diffused atoms in both structures to keep
the inner PMI shells occupied with 9~Al atoms, and repeated the
annealings. Then both structures turned out to be stable with respect to
  all competing phases. Thus, our final optimal models of the $\xi$-phase
contain 152~atoms in the primitive cell, with two stable compositions
Al$_{56}$Pd$_{17}$Mn$_{3}$ and Al$_{57}$Pd$_{16}$Mn$_{3}$, respectively. The
space group of both structures is $Cmc2_{1}$ (No.~36). 

\begin{figure}[!h]
  \centering
  \includegraphics[width=1\linewidth]{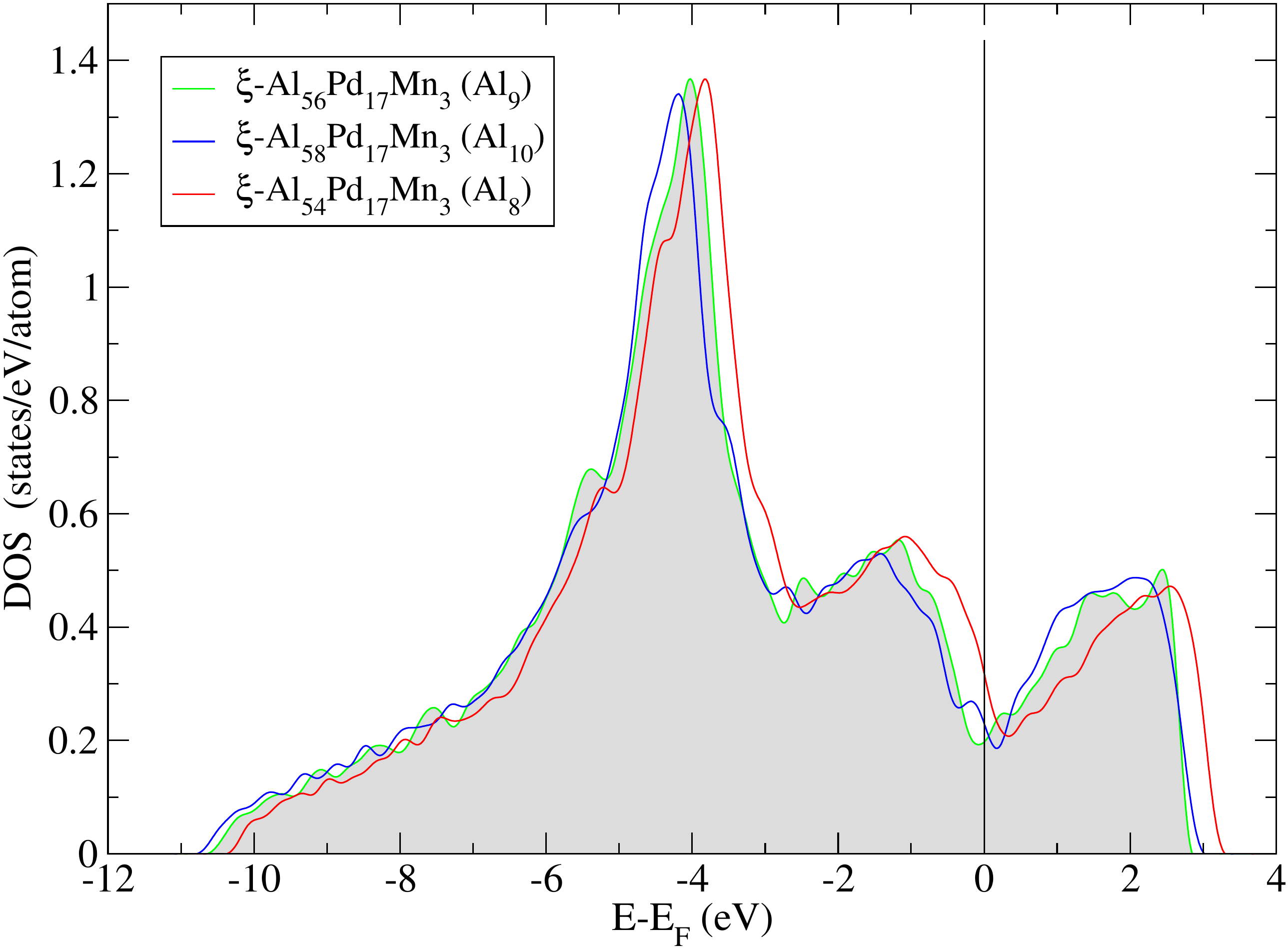}
  \caption{(Color online) Comparison of the electronic density of states (DOS)
    of $\xi$ with 
    different inner PMI shells. The structure with Al$_{9}$ inner shells is
    stable with respect to the competing phases, whereas the structures with
    Al$_{8}$ and Al$_{10}$ shells are unstable by 28.0 and 4.7~meV/atom,
    respectively.}
  \label{fig:xi-stable-eDOS}
\end{figure}

The electronic density of states of the former structure is shown in
Fig.~\ref{fig:xi-stable-eDOS}. Its prominent feature is the broad and deep
pseudogap. While models with Al$_9$ or Al$_{10}$ shells have Fermi energies
located near the center of the pseudogap, as required by the electronic
stabilization mechanisms, models with unfavorable inner shells, as for
instance Al$_8$, have Fermi energies shifted by about 0.3~eV relative to the
pseudogap minimum.

Finally, we have applied these optimal decoration rules to the $\xi'$-tiling,
and relaxed the models until forces vanished. The resulting total energies
were by 0.4~meV/atom lower than their $\xi$-tiling counterparts (for both stable
compositions). The space group of both structures is $Pna2_{1}$ (No.~33). 
In Ref.~\onlinecite{relaxed-structures} we list the relaxed atomic positions
of our stable $\xi$ and $\xi'$ structures.

\subsubsection{Significance of the Al(9) atoms}
\label{sec:cluster-free-atoms}
So far our models did not include the Al(9) atoms located outside the PMI and
LBPP clusters [see Fig.~\ref{fig:boudard_xipr}]. From the diffraction data
refinement, the site was half-occupied by Al, but its occupancy conflicted
with the pair of nearby (about 2.3~\AA) Al(17) atoms, that were reported to be
fully occupied. Nevertheless, for completeness we studied also various models
in which this site gets occupied by an Al atom (while the conflicting Al(17)
atoms were expected to adjust their position to avoid the short distance).

We observed that during the annealing these atoms drifted either into the PMI
clusters and increased the occupancies of the inner shells, or into the LBPP
clusters if there were any vacancies available.  Only when the annealing
temperature was kept as low as 300--500~K, the Al atoms remained at these
Al(9) sites. In our stable models we fully occupied these sites and relaxed
the coordinates until forces vanished. The structures were unstable by more
than 7~meV/atom, corresponding to about 0.5~eV cost per Al(9) site. Possibly,
these sites are a feature of the high-temperature structure.

\subsection{Internal Degrees of Freedom of Inner PMI Shells in
  \texorpdfstring{$\xi$ and $\xi'$}{xi and xi'}}
\label{sec:ab-intio-MD}
The Al$_9$ shells forming the core part of the PMI clusters have approximate 
trigonal 3/m symmetry, but they are encapsulated within an approximately
icosahedral potential of the second PMI shells Pd$_{12}$Al$_{30}$. Due to this
broken symmetry it is natural to expect that there must exist other,
energetically nearly equivalent orientations of the Al$_9$ shell. On the other
hand, there are two additional strong terms effectively lowering the symmetry
of the potential inside the PMI: approximately pentagonal symmetry breaking
due to having one 5-fold axis parallel to the global pseudo-decagonal (stacking)
axis, and the PMI--PMI direct linkages, dictated by the arrangement of the
PMI-cluster columns.


\begin{figure}[!h]
  \centering
  \includegraphics[width=1\linewidth]{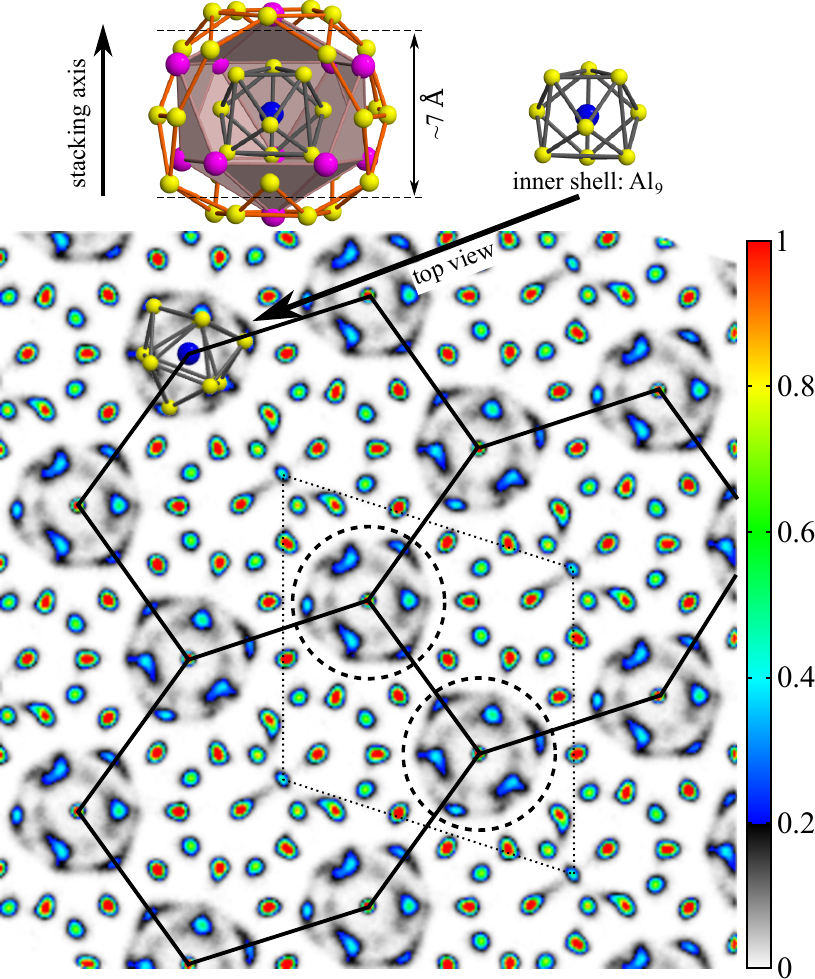}
  \caption{(Color online) Density plot of the atoms in the $\xi$-phase at
    1200~K, time 
    averaged after 50 ps. The density plot is projected along the PMI stacking
    axis. Dotted lines outline the unit cell, dashed circles outline the inner
    PMI shells. The upper part shows a single PMI cluster and its inner
    Al$_{9}$ shell: only atoms within the dashed lines are shown in the
    density plot. For the sake of clarity, one Al$_{9}$ shell is superimposed
    in the density plot.}
  \label{fig:xi-dynamics}
\end{figure}

The vibrational and configurational degrees of freedom related to the
reshufflings of the inner-shell atoms are activated at finite temperature, and
certainly contribute to the free energy, providing a stabilization effect. 

Here the dynamics of the inner shells has been studied in an {\it ab initio}
molecular dynamics simulation at 1200~K in the NVT ensemble. A time-averaged
density plot of a 7~\AA\ thick slice containing one ``layer'' 
of PMI clusters, projected along to the PMI stacking axis, is shown in
Fig.~\ref{fig:xi-dynamics}. The sharp red-yellow maxima correspond to the
rigid part of the structure, with atomic positions concentrated around their
unique equilibrium positions.  The weaker, smeared blue-gray areas correspond
to extremely mobile inner-shell atoms of the PMI clusters. While reshuffling,
the atoms approximately maintain the trigonal prismatic shape of the shell,
resulting in a kind of rotational motion: no distinct jumps could be observed.

\begin{figure}[h]
  \centering
  \begin{minipage}[c]{0.48\linewidth}
    \includegraphics[width=0.7\linewidth]{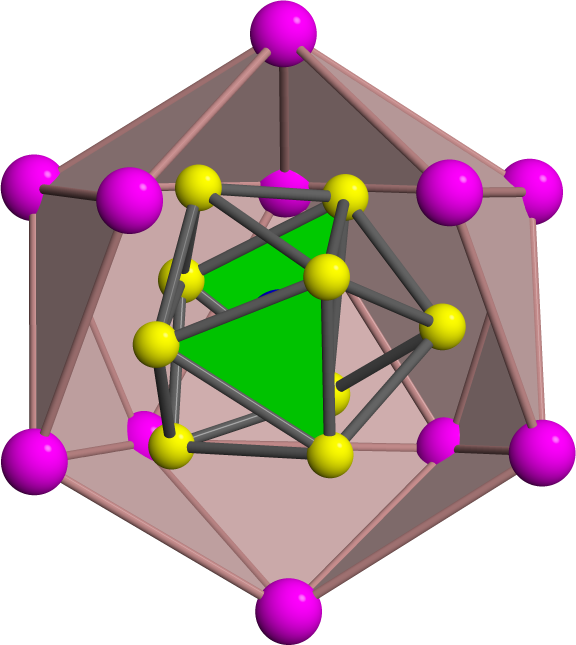}
  \end{minipage}
\begin{minipage}[c]{0.48\linewidth}
    \includegraphics[width=0.7\linewidth]{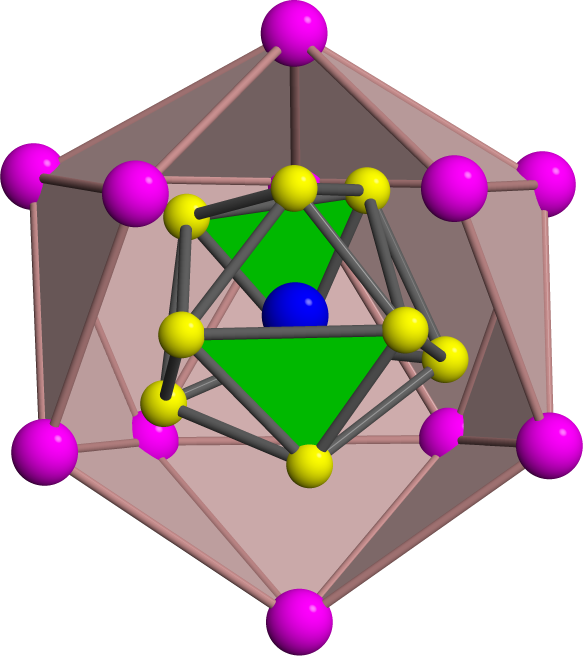} 
  \end{minipage} 
  \caption{(Color online) Snapshots of an inner PMI shell at different times
    after 
    relaxation. The vertical 5-fold axis is the stacking axis of the PMI
    clusters. The triangles of the trigonal prisms are highlighted in green to
    outline their 3-fold axis.}
  \label{fig:inner-shell-rotation}
\end{figure}

To get a first estimate for the energy fluctuations as the inner-shell atoms
reshuffle, we relaxed the high-temperature structure at different times to
zero Kelvin. The relaxed energies varied up to about 2~meV/atom. In all cases
the 3-fold axes of the inner Al$_{9}$ shell pointed along a local 5-fold
axis of the outer PMI shells (not necessarily the stacking axis). In addition,
we found certain configurations where the same inner shell appeared slightly
rotated with respect to the same 5-fold axis as shown in
Fig.~\ref{fig:inner-shell-rotation}.

Nevertheless, the time-averaged density plot indicates that the preferred
orientations of the inner shells must be parallel to the stacking axis of the
structure. In fact, in our lowest energy configurations the 3-fold axes of all
inner shells point along the stacking direction.

\section{Discussion}
\label{sec:discussion}

\subsection{\texorpdfstring{$\xi'$}{xi'} in the Binary Al--Pd System}
\label{sec:al3pd}
In 1994 Matsuo {\it et al.}\cite{matsuo_1994} reported on an approximant to
a decagonal quasicrystal with 16~\AA\ periodicity, designated Al$_{3}$Pd. The
orthorhombic unit cell contained approximately 280 atoms occupying 300 sites,
with lattice parameters $a=23.36$~\AA, $b=12.32$~\AA\ and $c=16.59$~\AA, and
space group $Pna2_{1}$. The structure is nearly identical to the $\xi'$-phase.
The main differences are the Mn sites that are substituted by Pd, and the
inner PMI shells, which contain at most 8~Al atoms. A direct comparison of the
atomic positions can be found in Ref.~\onlinecite{boudard_structure_1996}.

As in the ternary Al--Pd--Mn case, for the purpose of optimization of total
energies we studied the binary $\xi$-phase, rather than the $\xi'$-phase. 
Interestingly, the interstitial Al(9) site of the Boudard model was also found
in Al$_{3}$Pd (Al(21) in Ref.~\onlinecite{matsuo_1994}). The occupancy factor
was specified as 0.2. Similar to the ternary $\xi$- and $\xi'$-phase, we found
that these atoms are evidently disfavoured by our total energy calculations.

Varying the number of atoms in the inner PMI shells from 8 to 10, we find that
like in the ternary phase, the optimal shells are Al$_9$ with similar trigonal
prismatic arrangements. The second modification that lowered the energy by
3~meV/atom was a removal of four Pd atoms from four PMI second shells. 

\begin{figure}[!h]
  \centering
  \begin{minipage}[b]{0.54\linewidth}
    \centering
    \includegraphics[scale=0.15]{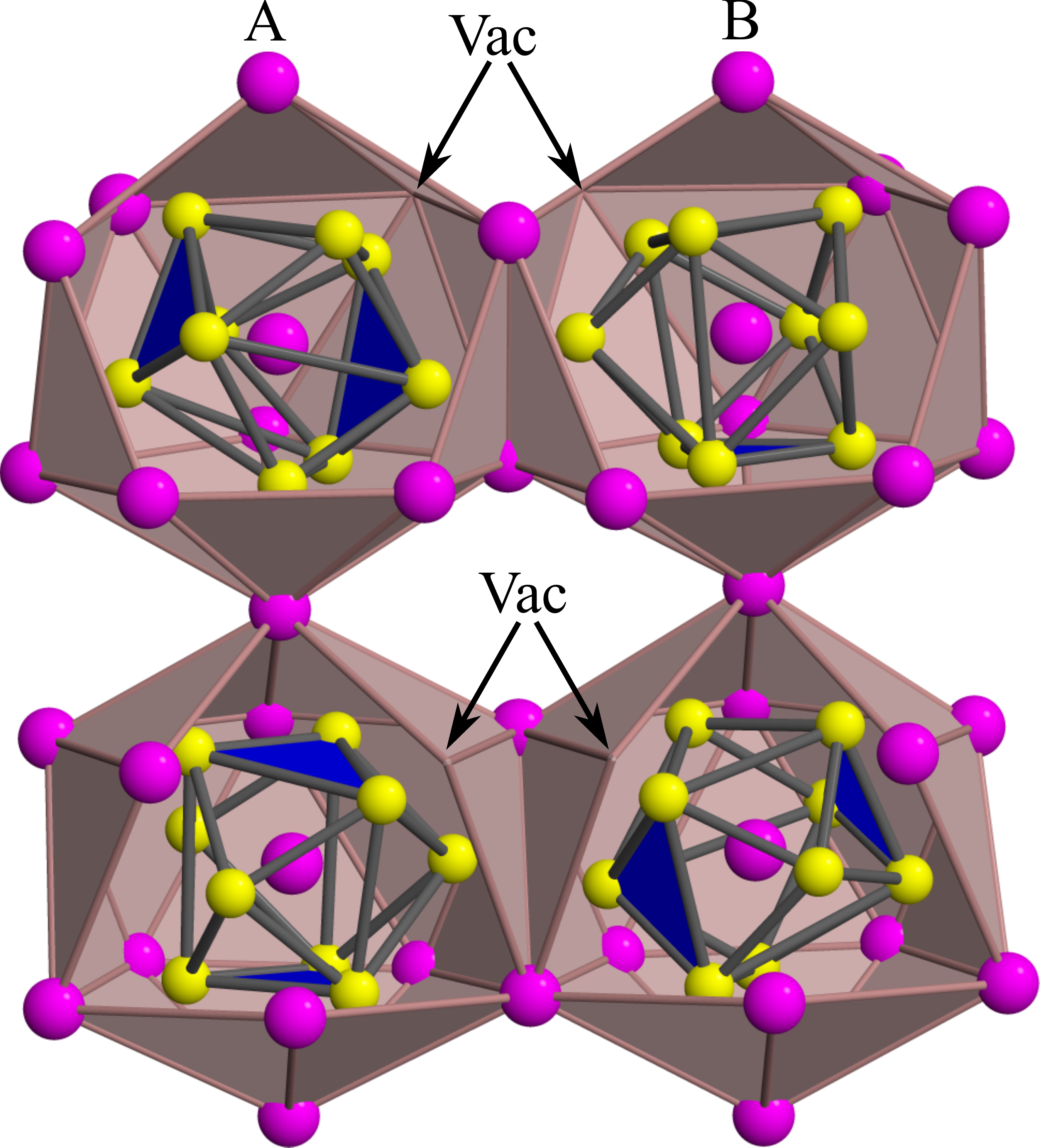}\\(a)\\
  \end{minipage}
  \begin{minipage}[b]{0.44\linewidth}
    \centering
    \includegraphics[width=1\linewidth]{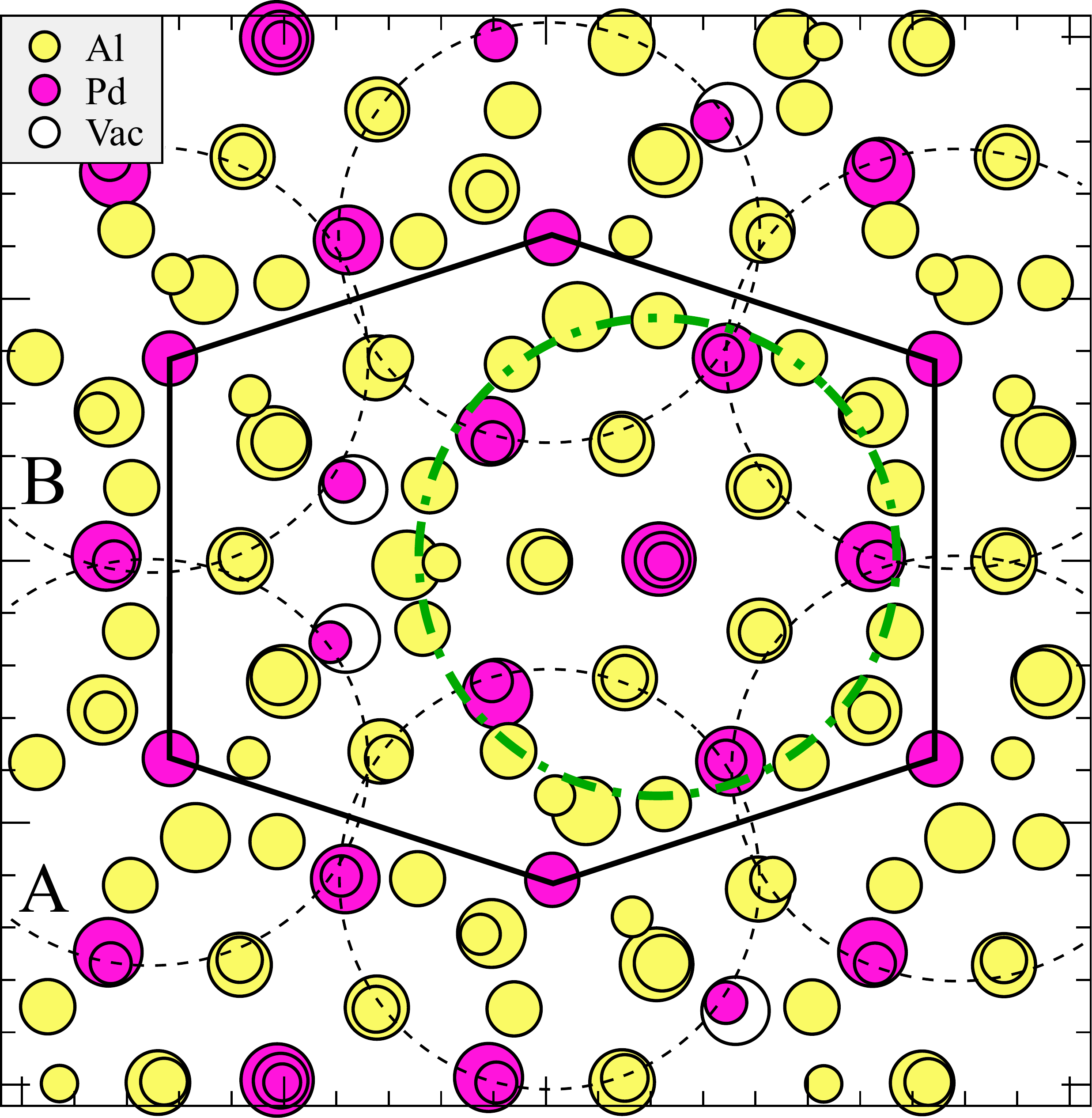}\\(b)\\
  \end{minipage}
  \caption{(Color online) Optimized Al$_3$Pd structure. 
    (a) Side view of PMI clusters with central Pd atoms, inner Al$_{9}$ and
    outer Pd$_{11}$ shells. (b) Top view along the stacking axis of PMI
    clusters. Only atoms on the flat and its adjacent puckered layers are
    shown. PMIs are encircled with black-dotted lines. Color coding same as in
    previous figures.}
  \label{fig:Al3Pd-structure}
\end{figure}

The four PMI clusters contained in the primitive unit cell of this binary
$\xi$-phase are shown in Fig.~\ref{fig:Al3Pd-structure}(a). 
The Pd~atoms were removed such that the vacancies are located next to each
other in the hexagons [cf.~Fig.~\ref{fig:Al3Pd-structure}(b)]. The centers of
the LBPP clusters and the centers of the icosahedra within the LBPPs are all
occupied by Pd: these three Pd atoms form a vertical chain with 2.86~\AA\
distance to each other. The unit cell contains 112~Al and 38~Pd atoms,
corresponding to a composition of about Al$_{74.7}$Pd$_{25.3}$. At zero
Kelvin, the structure is by 15~meV/atom unstable to decomposition into $fcc$
Al and Al$_{21}$Pd$_{8}$. At elevated temperatures, instead of $fcc$ Al,
the competing phase should be Al$_4$Pd. 
Interestingly, both of these phases are packings of Al$_9$Pd clusters, that
appear to be a distorted version of the Al$_9$Mn clusters found inside the PMI
clusters of the $\xi$-phase. The metastability of these binary $\xi$- and
$\xi'$-phases is in agreement with the experimental data.

In the optimal structure described above, some Al$_9$Pd clusters are rotated
such that their 3-fold axis is no longer parallel to the stacking direction of
the structure, which is a result of molecular dynamics annealing. If we
insisted they all aligned along the stacking axis, the structure is mechanically
stable, and its symmetry would be consistent with the experimentally
determined space group $Pna2_{1}$, but the energy would increase by another
2~meV/atom. Of course, this agreement should be interpreted as accidental:
even low-symmetry models may still belong to a finite-temperature ensemble
with higher statistical symmetry.

\subsection{Toward Phason-Line Models}
\label{sec:phason-lines}
The $\varepsilon$-phase family consists, apart from $\xi$ and $\xi'$, of a
variety of other similar phases. These phases can be described by introducing
an additional set of tiles, a banana-shaped nonagon and a pentagon. As before,
each vertex is a projection of the PMI clusters along their stacking
axis. The tiles are always found attached to each other, and have been termed
phason defect or phason line.\cite{Klein_1999} Periodic sequences
of these phason lines are called phason planes.\cite{klein_2003} The
``larger'' $\varepsilon$-phases are distinguished by the number of hexagon rows
that are sandwiched in between these phason planes. Their lattice parameters
along the {\bf a} and {\bf b} directions are the same, whereas the {\bf c}
parameter depends on the number of inserted hexagon
rows.\cite{heggen_structural_2008} The simplest of these phases is $\xi_{1}'$
($\varepsilon_{16}$) which contains exclusively phason planes with no hexagon
tiles. 

Only a few atomistic models exist in the literature describing these larger
phases. No experimental structure refinements have been carried out yet,
primarily due to the lack of good quality single crystals. Most investigations
rely on the average structure of $\xi'$. Tian {\it et al.} constructed a
$\Psi$-phase by means of a {\it cut-and-shift} method of
$\xi'$.\cite{tian_2006,tian_2006_2} Thereby, two parts of the same $\xi'$
structure are translated relative to each other. The model is confirmed by
comparing the experimental and simulated high-resolution transmission electron
microscopy images. More recently, a structure model for the
$\xi_{1}'$-phase in the Al--Rh system has been proposed using a 
{\it strong-reflections approach}.\cite{Mingrun_2010} Similarly to Tian 
{\it et al.}, the atomic positions are deduced from those of the known 
$\xi'$-phase. The space group of both models is $B2mm$ (No.~38).

\begin{figure}[!h]
  \centering
    \begin{minipage}[c]{0.4\linewidth}
    \centering
    \includegraphics[scale=0.4]{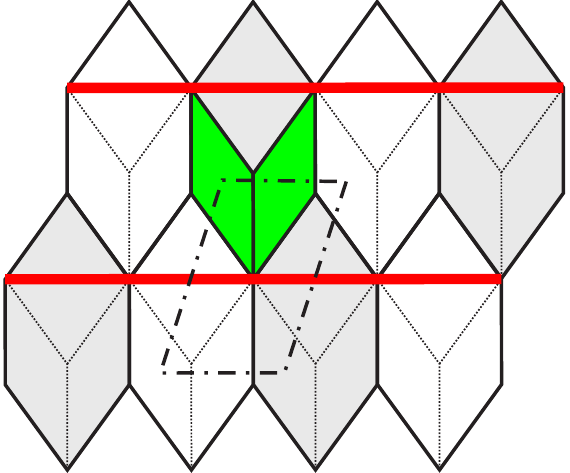}\\
    \centering $H$\\

    \vspace{0.2cm}
    \centering
    \includegraphics[scale=0.4]{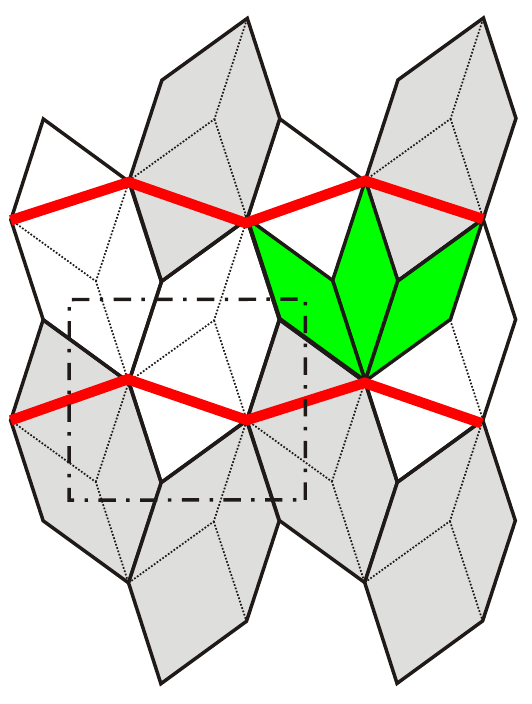}\\
    \centering $H'$
  \end{minipage}
  \begin{minipage}[c]{0.58\linewidth}
    \centering
    \includegraphics[scale=0.4]{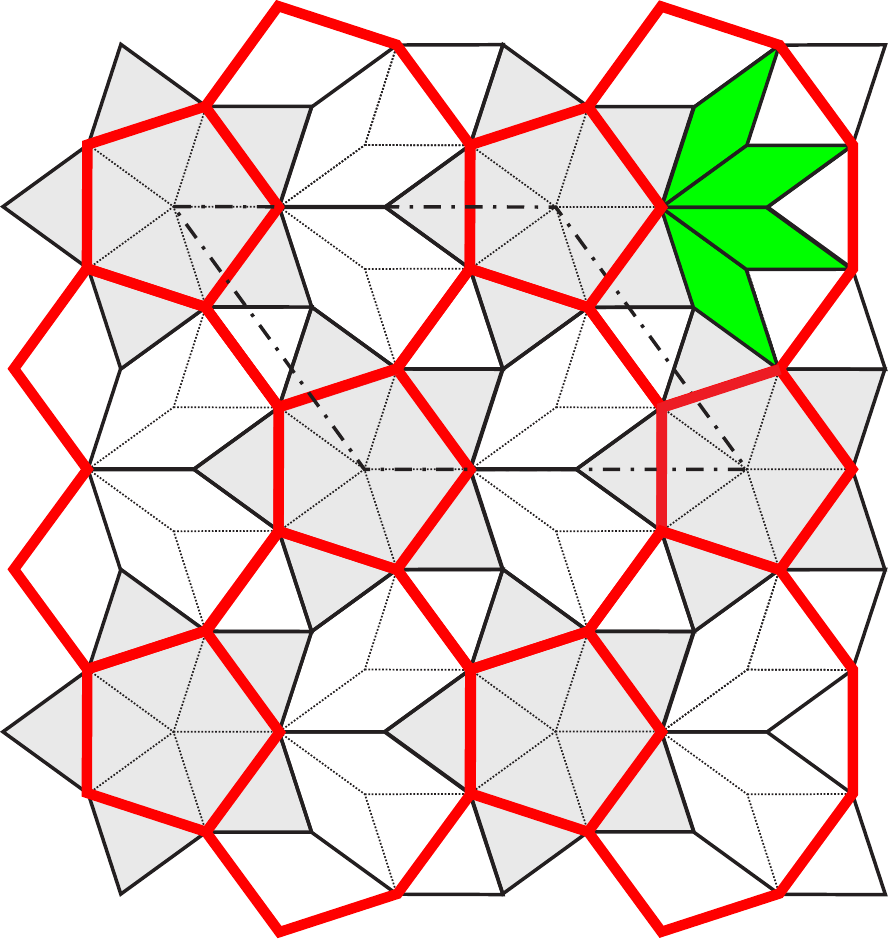}\\
    \centering $\xi_{1}'$
  \end{minipage}
  \caption{(Color online) HBS-tiling models for the $H$, $H'$ and
    $\xi_{1}'$-phases. The skinny 
    rhombi are outlined to illustrate the different tile-tile interactions
    existing in each phase. See text for more details.}
  \label{fig:hbstilings2}
\end{figure}

Here we use the $\xi_{1}'$-phase to obtain a microscopic model for the
phason lines. We construct the proper HBS tiling, and apply our optimized
decoration motifs obtained from the low-temperature $\xi$- and $\xi'$-phases.
In our HBS-tiling picture the pentagon is represented by a large Star tile,
and the nonagon by two Hexagons [see Fig.~\ref{fig:hbstilings2}].
Within the tiling the skinny Penrose rhombi are located next to each other. 
This results in a direct contact of the LBPP clusters in the structure, a
feature that is not present in $\xi$ or $\xi'$.

Before tackling the chain of four skinny rhombi in $\xi_1'$, we constructed
two other tilings in which the short diagonals of skinny rhombuses form
infinite chains [see Fig.~\ref{fig:hbstilings2}]. Both tilings are composed of
Hexagon tiles only. Due to their similar arrangement with the hexagons in
$\xi$ and $\xi'$, we termed these tilings $H$ and $H'$, respectively. The
structure of $H$ consists of PMI planes, formed by connecting adjacent
PMI-cluster columns. The LBPP clusters, on the other hand, form further planes
which are sandwiched in between the PMI planes. The primitive unit cell
contains approximately 100 atoms. Similarly, the $H'$-phase is simply an
arrangement of staggered PMI planes, with the LBPP clusters located in
between. The unit cell contains twice as many atoms as in the $H$-phase. Both
structures have not been observed in the experiment so far. Nevertheless, due
to their rather small unit cells they offer a fast way to calculate total
energies for a variety of decoration motifs. Afterwards, these motifs can be
applied to the $\xi_{1}'$-phase, and hence to every other $\varepsilon$-phase.

\begin{figure}[h]
  \centering
  \includegraphics[width=0.7\linewidth]{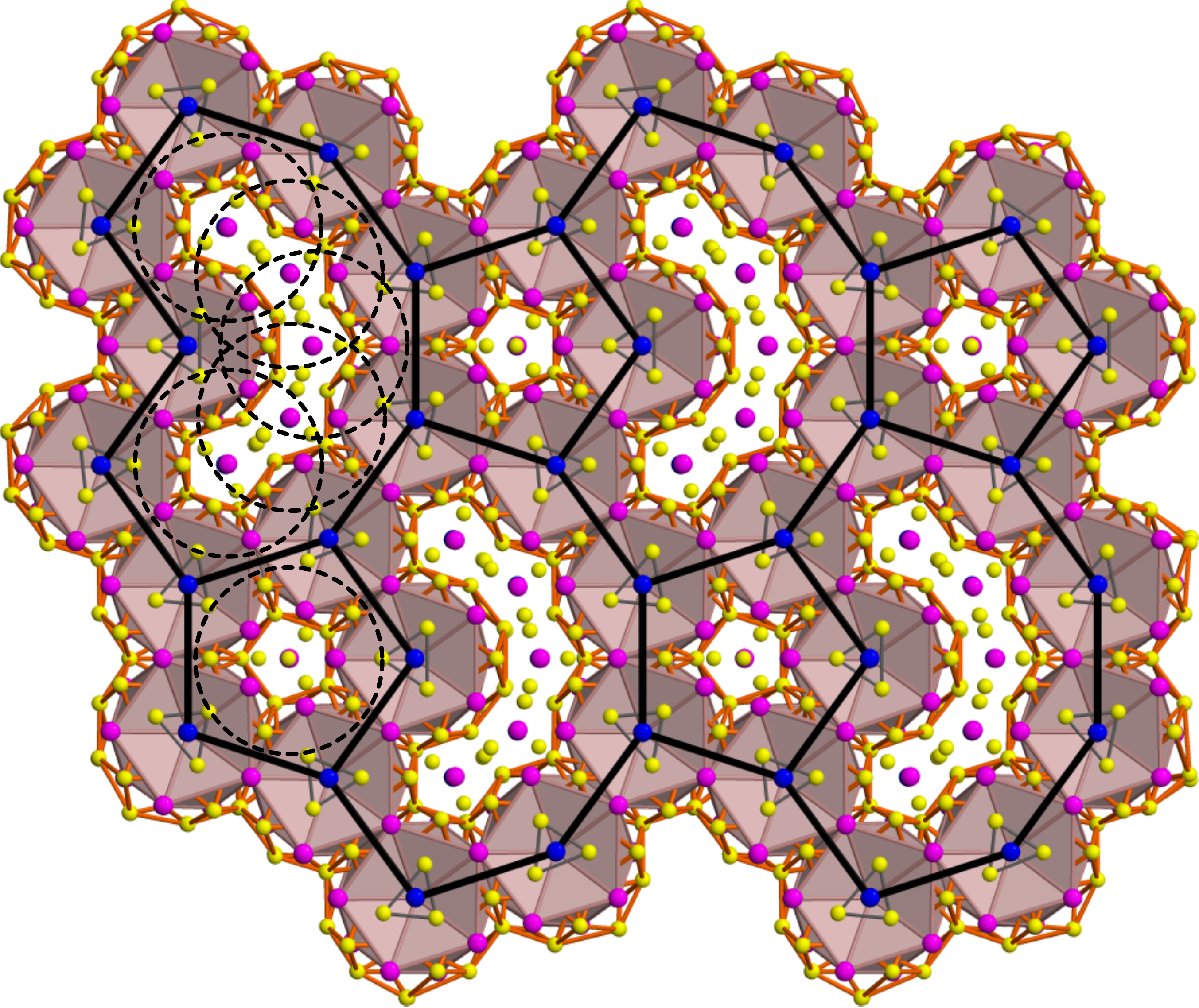}\\

  \vspace{-0.2cm}
  (a)\\

  \vspace{0.3cm}
  \includegraphics[width=0.8\linewidth]{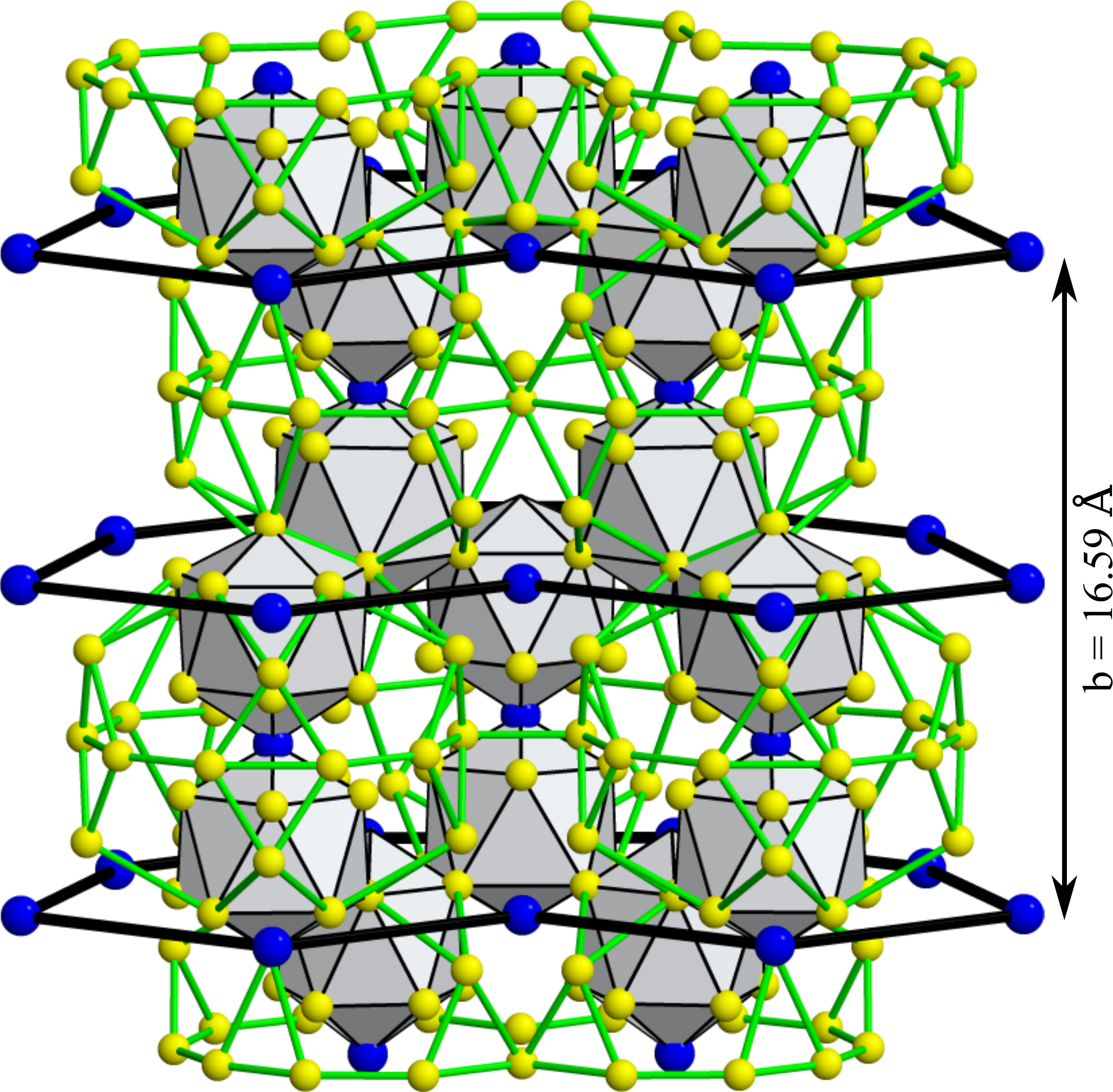}\\
  (b)\\
  \caption{(Color online) Optimized structure model for $\xi_{1}'$ with 398
    atoms in the primitive unit cell. (a) Projection of the structure
   along the PMI stacking axis. Dashed circled outline the locations of the
   LBPP clusters. (b) Side view of one nonagon tile. The Pd$_{10}$
   pentagonal prisms and the PMI-cluster columns are omitted for the sake of
   simplicity. The nonagon contains five LBPP clusters per unit-cell height.}
  \label{fig:xi1prime-optimal}
\end{figure}

Our lowest-energy structure of the $\xi_{1}'$-phase contains nearly the
same (symmetry-broken) LBPP clusters as our optimized $\xi$ and $\xi'$ models
[cf.~Fig.~\ref{fig:LBPPs}(b)]. The inner PMI shells are all occupied by 9 Al
atoms, forming the usual trigonal-prismatic shells. The banana-shaped nonagon
comprises five entire LBPP clusters per unit cell height, as shown in
Fig.~\ref{fig:xi1prime-optimal}. The nonagon tile, on
the other hand, contains only one LBPP cluster per unit cell height. The
structure is about 2.5~meV/atom above our convex hull. The primitive unit cell
contains 294~Al, 88~Pd, and 16~Mn atoms. The space group is $C1m1$ (No.~8). 
The relaxed atomic positions of this structure can be found in
Ref.~\onlinecite{relaxed-structures}.

Our highest-symmetry model for the $\xi_{1}'$-phase is unstable by
11~meV/atom. The space group is in agreement with the experimentally found
ones. The structure is in principle the same as our best model, except
for the LBPP clusters: here the Al atoms along the LBPP 5-fold axis are
substituted by Pd atoms [cf.~Fig.~\ref{fig:LBPPs}(a)].

For comparison, we also relaxed the structure model for the $\Psi$-phase,
proposed by Tian {\it et al.}, with 540~Al, 159~Pd, and 36~Mn atoms in the
primitive cell. According to our DFT calculations, their model is
unstable by more than 30~meV/atom with respect to our competing phases. This
rather high energy difference is not surprising: in contrast to our models,
their PMI clusters contain Al$_{10}$ shells, as well as several (unfavored)
interstitial atoms that are not covered by any cluster. The chemical
orderings inside their LBPP clusters are also in contradiction to our
optimized models. Furthermore, their banana-shaped nonagons contain only four
LBPP clusters per unit cell height.

With the refined models for the flattened hexagons in $\xi$ and $\xi'$,
and the optimized models for the phason lines, we are now able to construct
all variants of the $\varepsilon$-phases family in the Al--Pd--Mn system. We
believe that knowing the optimal decoration motifs for the fundamental tiles
enables us to construct all other possible structures by simply changing the
underlying tiling in our tiling-decoration model.

\begin{figure*}
  \centering
  \begin{minipage}[b]{0.49\linewidth}
    \centering
    \fbox{\includegraphics[width=0.95\linewidth]{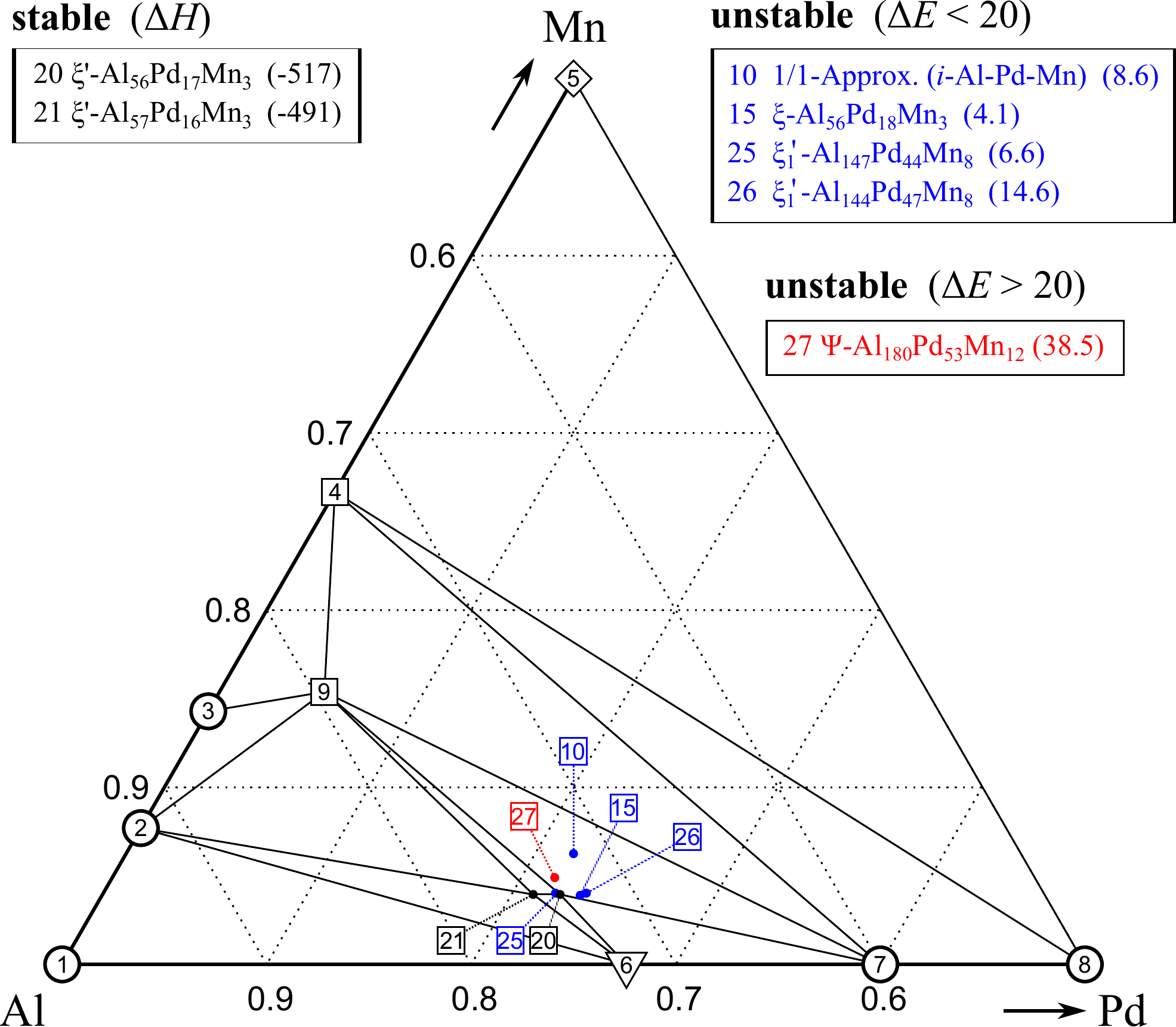}}
  \end{minipage}
  \begin{minipage}[b]{0.49\linewidth}
    \centering
    \fbox{\includegraphics[width=0.935\linewidth]{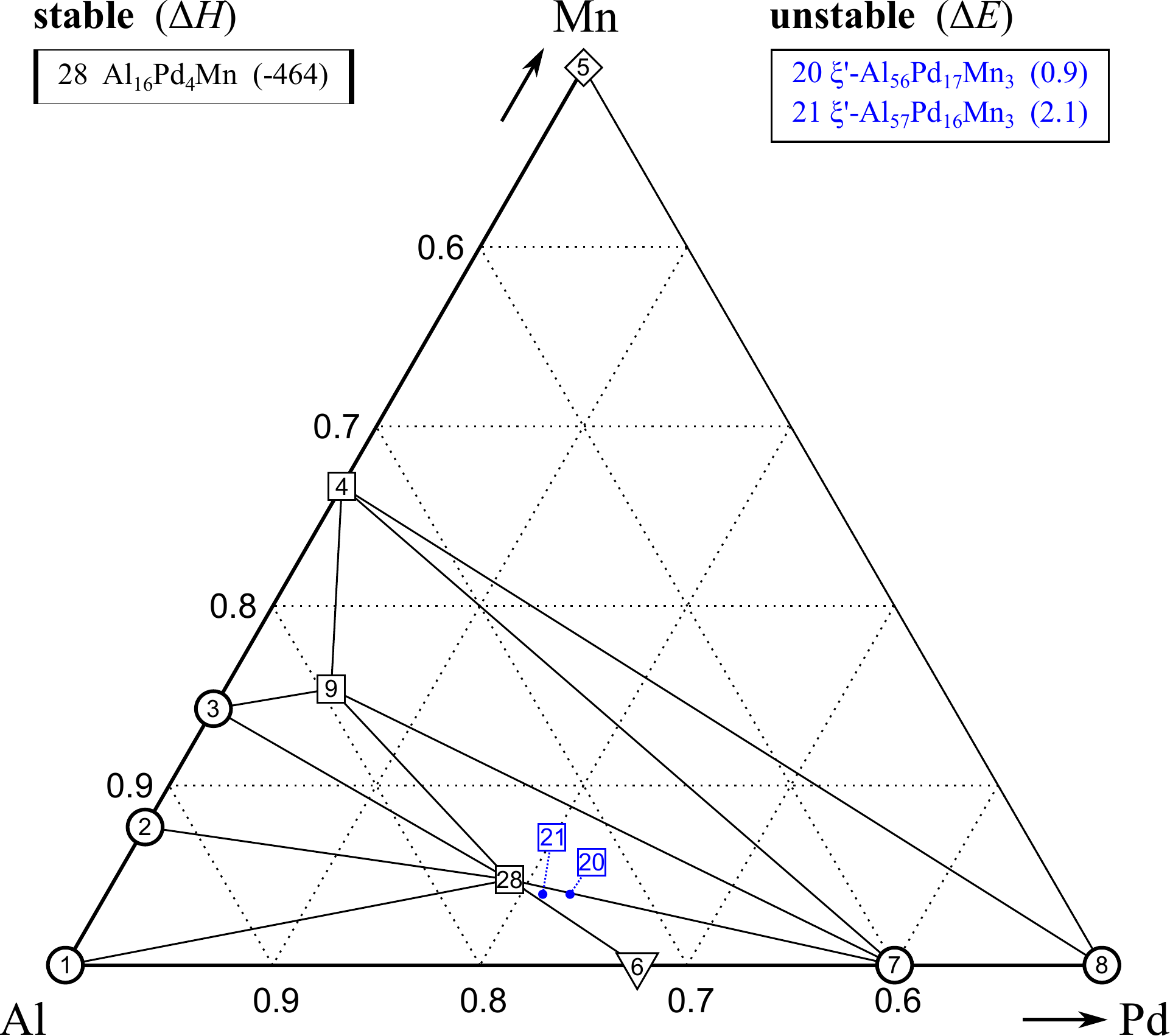}}
  \end{minipage}
  \caption{(Color online) Low-temperature energy-phase diagrams of the
    Al--Pd--Mn system in the 
    Al-rich corner with our optimized structures. The structures are labeled
    according to Tab.~\ref{tab:competing-phases} and
    \ref{tab:low-temp-xiphases}. Left: The two $\xi'$-phases are stable down
    to $T$=0~K with respect to all competing phases. The recently discovered
    Al$_{16}$Pd$_{4}$Mn phase is not included in the stability evaluation. 
    Right: If we include the Al$_{16}$Pd$_{4}$Mn phase in our stability
    evaluations, both $\xi'$-phases lie slightly above the convex hull.}
  \label{fig:phase-diagram-new}
\end{figure*}

Tab.~\ref{tab:low-temp-xiphases} summarizes all optimized structures discussed
in this work by listing compositions, space groups, Pearson symbols and
energies. The structures are shown in the same order as appearing in
Sec.~\ref{sec:Optimization} and \ref{sec:discussion}. Our new low-temperature
energy-phase diagrams with the optimized structures are shown in
Fig.~\ref{fig:phase-diagram-new}. For completeness, we also include the
recently discovered Al$_{72}$Pd$_{18}$Mn$_{5}$Si$_{5}$ phase in a separate
phase diagram, with Si substituted by Al yielding a composition of
Al$_{16}$Pd$_{4}$Mn. Surprisingly, the structure is stable down to $T$=0~K, 
with respect to all competing phases including our optimized $\xi'$-phases
[see right part of Fig.~\ref{fig:phase-diagram-new}]. The calculated relative
stability of this phase at $T$=0~K does not change any of the conclusions
about the optimal structure of the $\varepsilon$-phases. It should also be
remembered that our exploration of the low-temperature phase stability remains
incomplete due to the absence of accurate models of the icosahedral Al--Pd--Mn
phase. 

\begin{table*}
  \caption{Compositions, space groups, Pearson symbols and energies of our
    optimized structures. The structures are listed in the same order as
    discussed in the paper. $\Delta E_{\text{old}}$ is the energy difference
    to our initial convex hull (cf.~Fig.~\ref{fig:alpdmn_phase_diagram}),
    $\Delta E_{\text{new}}$ is the energy difference relative to our new convex
    hull (cf. left part of Fig.~\ref{fig:phase-diagram-new}). The relaxed
    atomic positions of structures 16, 17, 20, 21 and 25 are listed in
    Ref.~\onlinecite{relaxed-structures}.}
\begin{ruledtabular}
  \begin{tabular}{rlcccddd}
    & & & & & \multicolumn{1}{c}{\textrm{$\Delta E_{\text{old}}$}} &
    \multicolumn{1}{c}{\textrm{$\Delta E_{\text{new}}$}} &
    \multicolumn{1}{c}{\textrm{Total Energy}} \\
    \# & Structure & PMI Shell & Space Group (No.) & Pearson Symbol &
    \multicolumn{1}{c}{\textrm{(meV/atom)}} &
    \multicolumn{1}{c}{\textrm{(meV/atom)}} &
    \multicolumn{1}{c}{\textrm{(eV/atom)}}\\
    \hline
    11 & $\xi$-Al$_{55}$Pd$_{16}$Mn$_{4}$ & Al$_{8}$ & $P1$ (1) & $aP150$ &
    42.5 & 45.8 & -4.7523 \\
    12 & $\xi$-Al$_{57}$Pd$_{16}$Mn$_{4}$ & Al$_{9}$ & $P\bar{1}$ (2) &
    $aP154$ & 13.9 & 17.6 & -4.7537 \\
    13 & $\xi$-Al$_{59}$Pd$_{16}$Mn$_{4}$ & Al$_{10}$ & $C12/c1$ (15) &
    $mC316$ & 12.6 & 16.0 & -4.7293 \\ 
    14 & $\xi$-Al$_{61}$Pd$_{16}$Mn$_{4}$ & Al$_{11}$ & $C1c1$ (9) & $mC324$ &
    33.6 & 37.4 & -4.6828 \\ 
    15 & $\xi$-Al$_{56}$Pd$_{18}$Mn$_{3}$ & Al$_{9}$ & $Cmcm$ (63) & $oC308$ &
    0.2 & 4.1 & -4.7741 \\
    16 & $\xi$-Al$_{56}$Pd$_{17}$Mn$_{3}$ & Al$_{9}$ & $Cmc2_{1}$ (36) & 
    $oC304$ & -3.9 & 0.4 & -4.7476 \\ 
    17 & $\xi$-Al$_{57}$Pd$_{16}$Mn$_{3}$ & Al$_{9}$ & $Cmc2_{1}$ (36) &
    $oC304$ & -3.8 & 0.4 & -4.7023 \\ 
    18 & $\xi$-Al$_{54}$Pd$_{17}$Mn$_{3}$ & Al$_{8}$ & $P1$ (1) & $aP148$ &
    28.0 & 31.9 & -4.7426 \\ 
    19 & $\xi$-Al$_{58}$Pd$_{17}$Mn$_{3}$ & Al$_{10}$ & $P1$ (1) & $aP156$ &
    4.7 & 8.5 & -4.7133 \\
    20 & $\xi'$-Al$_{56}$Pd$_{17}$Mn$_{3}$ & Al$_{9}$ & $Pna2_{1}$ (33) & 
    $oP304$ & -4.3 & 0 & -4.7480 \\ 
    21 & $\xi'$-Al$_{57}$Pd$_{16}$Mn$_{3}$ & Al$_{9}$ & $Pna2_{1}$ (33) &
    $oP304$ & -4.2 & 0 & -4.7027 \\
    22 & $\xi$-Al$_{57}$Pd$_{17}$Mn$_{3}$ & Al$_{9}$ & $Cmc2_{1}$ (36) &
    $oC308$ & 7.0 & 10.9 & -4.7238 \\
    23 & $\xi$-Al$_{58}$Pd$_{16}$Mn$_{3}$ & Al$_{9}$ & $Cmc2_{1}$ (36) &
    $oC308$ & 7.9 & 11.8 & -4.6778 \\
    24 & $\xi'$-Al$_{56}$Pd$_{19}$ & Al$_{9}$ & $Pna2_{1}$ (33) & $oP300$ &
    16.6 & 16.6\footnotemark[1] &  -4.5772 \\
    25 & $\xi_{1}'$-Al$_{147}$Pd$_{44}$Mn$_{8}$ & Al$_{9}$ & $C1m1$ (8) & 
    $mC796$ & 2.5 & 6.6 & -4.7371 \\
    26 & $\xi_{1}'$-Al$_{144}$Pd$_{47}$Mn$_{8}$ & Al$_{9}$ & $Amm2$ (38) &
    $oA796$ & 11.0 & 14.6 & -4.7791 \\
    27 & $\Psi$-Al$_{180}$Pd$_{53}$Mn$_{12}$ & Al$_{10}$ & $Amm2$ (38) &
    $oA1470$ & 35.0 & 38.5 & -4.7429 \\
    28 & Al$_{16}$Pd$_{4}$Mn & Al$_{10}$ & $Pnma$ (62) & $oP168$ & -8.5 & -4.7
    & -4.6890
  \end{tabular}
\end{ruledtabular}
\footnotetext[1]{The discovery of a new stable ternary structure does not
  affect the stability of a binary structure.}
\label{tab:low-temp-xiphases}
\end{table*}

\section{Summary}
\label{sec:summary}
We have systematically optimized the structures of the $\xi$- and
$\xi'$-phases, and established all important correlations in chemical ordering
and occupancies. The structure of these phases is best described as an assembly
of two types of overlapping clusters, namely the pseudo-Mackay icosahedron and
the so-called ``large bicapped pentagonal prism''. The fundamental clusters 
comprising almost 90\% of all atoms are pseudo-Mackay icosahedra; they differ
from the well known Mackay icosahedron cluster by a reduction of the occupancy
and the symmetry of the inner shell, from 12 Al atoms (on the vertices of an
icosahedron) to only 9 Al atoms forming trigonal-prismatic shells. For the
secondary cluster we have determined the chemical ordering along its
pentagonal axis which minimizes the energy of the structure. We found that the
$\xi$- and $\xi'$-phases are, according to our calculations, stable down to
$T$=0~K (with a tiny preference for $\xi'$), and might even be a line compound,
since we found two discrete stable compositions:
Al$_{56}$Pd$_{17}$Mn$_{3}$ and Al$_{57}$Pd$_{16}$Mn$_{3}$. Both unit cells
contain 152~atoms (corresponding to 304 in $\xi'$). In an {\it ab initio}
molecular dynamics simulation we showed that the Al atoms of the Al$_{9}$ shells
inside the PMI clusters are highly mobile which can lead to partial occupancy
factors as obtained in the X-ray refinement.

A comparison has been made with the nearly isostructural Al$_{3}$Pd phase. The
structure can be described by the same type of clusters. However, in contrast
to the ternary phases, only metastable structures were found.

Using a tiling-decoration method we applied our optimized decoration motifs
obtained from the refinement of the $\xi$- and $\xi'$-phases to a tiling
representing the $\varepsilon_{16}$-phase. This phase enabled us to investigate
the structure of the phason defects in more detail. With these results we are
now able to construct all variants of the $\varepsilon$-Al--Pd--Mn phases. 

\vskip 1cm

\begin{acknowledgments}
This work has been supported by Deutsche Forschungsgemeinschaft, Paketantrag
``Physical Properties of Complex Metallic Alloys'' (PAK 36), TR-154/24-2. 
M.M. has been also supported by Slovak national grants VEGA j2/0111/11 and
APVV-0647-10.
\end{acknowledgments}



%
\end{document}